\documentclass[prd,preprint,tightenlines,showpacs,preprintnumbers,nofootinbib,eqsecnum,superscriptaddress,longbibliography,a4paper]{revtex4-1}

\usepackage[dvips,final]{graphicx}
\usepackage{amssymb}
\usepackage{amsmath}
\usepackage{amsfonts}
\usepackage{epsfig}
\usepackage{bm}

\usepackage[colorlinks=true, allcolors=blue]{hyperref}

\usepackage[section]{placeins}

\usepackage{sidecap}
\usepackage{multirow}
\usepackage{booktabs}
\usepackage{array}
\usepackage{tabularx}
\usepackage{xcolor}
\usepackage{pstricks}
\usepackage{color}
\usepackage{diagbox}
\usepackage{comment}
\usepackage{slashed}

\usepackage{subfig}
\usepackage{float}
\usepackage{caption}

\allowdisplaybreaks[4]

\newcommand{\ket}[1]{{#1} \rangle}
\newcommand{\bra}[1]{\langle {#1} }

\newcommand{\Bigket}[1]{{#1} \Big\rangle}
\newcommand{\Bigbra}[1]{\Big\langle {#1} }

\newcommand{\Ree}[1]{\text{Re}{#1}}
\newcommand{\Ime}[1]{\text{Im}{#1}}

\begin{document}

\title{Transition Form-Factor for $\eta_Q$ at NNLO in the strong coupling $\alpha_s$ and with all-order $v^2$ resummation}

\author{Izabela Babiarz}
\email{izabela.babiarz@ifj.edu.pl}
\affiliation{Institute of Nuclear Physics Polish Academy of Sciences, 
ul. Radzikowskiego 152, PL-31-342 Krak{\'o}w, Poland}

\author{Chris~A.~Flett}
\email{christopher.flett@ijclab.in2p3.fr}
\affiliation{Universit\'e Paris-Saclay, CNRS, IJCLab, 91405 Orsay, France}

\author{Melih~A.~Ozcelik}
\email{melih.ozcelik@ijclab.in2p3.fr}
\affiliation{Universit\'e Paris-Saclay, CNRS, IJCLab, 91405 Orsay, France}

\author{Wolfgang Sch\"afer}
\email{wolfgang.schafer@ifj.edu.pl} 
\affiliation{Institute of Nuclear
Physics Polish Academy of Sciences, ul. Radzikowskiego 152, PL-31-342 
Krak{\'o}w, Poland}

\author{Antoni Szczurek}
\email{antoni.szczurek@ifj.edu.pl}
\affiliation{Institute of Nuclear
Physics Polish Academy of Sciences, ul. Radzikowskiego 152, PL-31-342 
Krak{\'o}w, Poland}
\affiliation{College of Mathematics and Natural Sciences,
University of Rzesz\'ow, ul. Pigonia 1, PL-35-310 Rzesz\'ow, Poland\vspace{5mm}}

\begin{abstract}

In this work, we discuss both relativistic and perturbative QCD corrections to the transition form-factor ${\cal F}_{\eta_Q}{(t_1,t_2)}$ for the process $\gamma^*(q_1) \gamma^*(q_2) \leftrightarrow \eta_Q(P)$ with dependencies on the normalised photon virtualities $t_1=q_1^2/m_Q^2$ and $t_2=q_2^2/m_Q^2$, where $m_Q$ is the heavy quark mass. We resum a class of relativistic corrections to all orders in the relativistic parameter $\langle v^2 \rangle_{\eta_Q}$. In addition, we include perturbative QCD corrections up to Next-to-Next-to-Leading Order (NNLO) in the strong coupling constant $\alpha_s$. This involves the computation of two-loop amplitudes with two off-shell photons. We explore three different phenomenological applications of our transition form-factors for the charmonium case. We first study the ratio $\vert {\cal F}_{\eta_c}{(t_1,0)} \vert/ \vert {\cal F}_{\eta_c}{(0,0)} \vert$ for the single space-like photon case and compare our results with the existing $\eta_c$ measurements from the \textsc{Ba}\textsc{Bar} collaboration. Secondly, we consider observables for the case of double space-like photons and discuss the impact of the different corrections. The NNLO corrections for this case are presented here for the first time in the literature. Finally, we revisit the decay width $\Gamma[\eta_c \rightarrow \gamma \gamma]$ and compare it with the existing PDG value.

\end{abstract}

\maketitle

\section{Introduction}
\label{sec:introduction}

The study of quarkonia, which are bound states consisting of a heavy quark and its anti-particle, provides a valuable probe of QCD in both the perturbative and the non-perturbative regimes. Charmonium ($c\bar c$) and bottomonium ($b\bar b$) states are routinely measured at different collider experiments. In this work, we focus on the pseudo-scalar quarkonium, $\eta_Q$, which is a spin-singlet $S$-wave bound state. It is represented by the $Q\bar Q$ pair in the $^1S_0$ configuration with the quantum numbers $J^{PC} = 0^{-+}$, where $J$ is the total angular momentum of the $Q\bar Q$ pair, $P$ is the spatial parity and $C$ is the charge-conjugation parity of the meson. For reviews on quarkonium phenomenology and prospects, we guide the reader to refs.~\cite{QuarkoniumWorkingGroup:2004kpm,Brambilla:2010cs,Chapon:2020heu,Boer:2024ylx}.

The $\eta_Q(1S)$ meson is in fact the ground state in the charmonium and bottomonium spectroscopy, while also their excited states $\eta_Q(nS)$ have been observed \cite{ParticleDataGroup:2024cfk}. Due to its positive charge conjugation, it can couple to only an even number of photons. As such, the exclusive $\eta_Q$ production in lepton-lepton collisions, $e^+{(p_1)} e^-{(p_2)} \rightarrow e^+{(p'_1)} e^-{(p'_2)}\, \eta_Q{(P)}$, proceeds predominantly via the fusion of two off-shell photons $\gamma^*{(q_1)} \gamma^*{(q_2)} \rightarrow \eta_Q{(P)}$. In the case where only one of the scattered leptons is tagged, one of the photons becomes quasi on-shell. The $\eta_c$ cross-section with dependence on one photon virtuality has been measured by the \textsc{Ba}\textsc{Bar} collaboration in ref.~\cite{BaBar:2010siw} which has received considerable attention in recent years.

Most theory predictions to describe the data are currently based on Non-Relativistic QCD (NRQCD) \cite{Bodwin:1994jh}. Within this framework, one can arrange the different energy scales in the quarkonium system as follows $m_Q > m_Q v > m_Q v^2 > \Lambda_{\rm QCD}$, where $m_Q$ is the mass of the heavy quarks, $m_Q v$ is its relative three-momentum, and $m_Q v^2$ is the kinetic energy in the rest frame of the quarkonium. As such, the amplitude, to describe the production or decay of quarkonia, admits an expansion in both the relativistic parameter $v^2$ and the perturbative QCD parameter $\alpha_s$. Both parameters can be of similar size, hence it is vital to take into account both relativistic and QCD corrections.

In this work, we compute the transition form-factor for the process, $\gamma^*{(q_1)} \gamma^*{(q_2)} \rightarrow \eta_Q{(P)}$, as a function of the normalised photon virtualities $t_1=q_1^2/m_Q^2$ and $t_2=q_2^2/m_Q^2$ at Next-to-Next-to-Leading Order (NNLO) in $\alpha_s$ and, in addition, we consider a class of relativistic corrections that we resum up to all orders in $v^2$. We focus on the kinematic regions where both photons become space-like with $t_1 \leq 0$ and $t_2 \leq 0$. Our results can be used to describe exclusive $\eta_Q$ production in lepton-lepton collisions, where both outgoing leptons are tagged. In the limit where one of the photons becomes on-shell, we can compare our results with the \textsc{Ba}\textsc{Bar} measurements for the charmonium case and, in the double on-shell case, they can be used to describe the decay width $\Gamma[\eta_Q \rightarrow \gamma \gamma]$.

On the perturbative QCD side, this involves the computation of two-loop diagrams with massive legs and multiple scales. The evaluation of such Feynman integrals can be quite challenging due to their mathematical and singularity structure \cite{Travaglini:2022uwo,Abreu:2022mfk,Weinzierl:2022eaz}. In ref.~\cite{Abreu:2022vei}, for the special case of the double on-shell limit $t_1=t_2=0$, the two-loop master integrals were computed fully analytically and used in two-loop form-factor calculations for the $^1S_0$ and $^3P_J$ states \cite{Abreu:2022cco,Ozcelik:2025uln}. In the case of one off-shell leg, differential equations for the master integrals were studied in ref.~\cite{Chen:2017soz,Chen:2017xqd}. For the process with two off-shell photon virtualities, the two-loop amplitudes and master integrals are unknown in the literature and, due to their increased complexity on the analytic side, we compute them numerically here.

In the literature, we note that almost all transition form-factor calculations have been performed with either only one off-shell leg or in the double on-shell limit. An early calculation of the single off-shell transition form-factor at Next-to-Leading Order (NLO) can be found in ref.~\cite{Shifman:1980dk}, while the NNLO calculation was done in ref.~\cite{Feng:2015uha}. These works, however, concentrated on the lowest order in the $v^2$ expansion, thus neglecting relativistic corrections. Such corrections were studied in the context of $\eta_Q$ decay to di-photon up to higher orders in the relativistic parameter $v^2$ in refs.~\cite{Bodwin:2002cfe,Brambilla:2006ph}. These analyses have been further complemented by mixed $\alpha_s$ and $v^2$ corrections for the same observable in refs.~\cite{Jia:2011ah,Guo:2011tz}.\footnote{We note also ref.~\cite{Sang:2009jc, Xu:2014zra, Chen:2017pyi} for calculations in the time-like region relevant for exclusive $\eta_Q+\gamma$ production.} Relativistic effects have also been explored using a variety of other approaches, such as lattice QCD \cite{Dudek:2006ut}, sum-rules \cite{Lucha:2012ar} or light-front wavefunctions \cite{Babiarz:2019sfa,Li:2021ejv}. Here, the calculations in refs.~\cite{Dudek:2006ut,Lucha:2012ar,Babiarz:2019sfa} also present results for the form-factor for two space-like photons.

In Section~\ref{sec:Rel_Corr}, we outline the NRQCD formalism for the transition form-factor and introduce the relativistic corrections. In Section~\ref{Sec:NNLO}, we then present for the first time in the literature the NNLO corrections in $\alpha_s$ for the case of two space-like photons. In Section~\ref{Sec:Applications}, we proceed to discuss three phenomenological applications of the transition form-factor for the charmonium case. We first study the single off-shell case and compare our results with the \textsc{Ba}\textsc{Bar} data. We then analyse the results for the double off-shell case, before revisiting the decay width to di-photon and exploring the impact of the different corrections. We conclude in Section~\ref{Sec:Conclusions}.

\section{Relativistic corrections to $^1S_0$ quarkonium}
\label{sec:Rel_Corr}

To understand quarkonium production cross-sections and decay, one can work within a comprehensive framework that involves expanding in both the QCD strong coupling $\alpha_s$ and the relative velocity parameter $v^2$ between the heavy quark and anti-quark quarkonium constituents. The two expansions inherently capture different phenomena and are systematically encoded within Non-Relativistic QCD~(NRQCD) \cite{Bodwin:1994jh}. This is a non-relativistic effective field theory that treats quarkonium states as non-relativistic systems while accounting for relativistic effects via higher-order corrections in the parameter $v^2$.

We consider the process $\gamma^*(q_1) \gamma^*(q_2) \to \eta_Q(P)$ for two off-shell photons with momenta $q_1$, $q_2$, and the pseudo-scalar $Q \bar Q$ bound state $\eta_Q$ with momentum $P$.
This amplitude can be parametrised in terms of one invariant transition form-factor as
\begin{equation}
 {\cal M}(\gamma^*(q_1) \gamma^*(q_2) \to \eta_Q(P)) = 
4 \pi \alpha_{\rm em} \, i \, \varepsilon_{\mu \nu \alpha \beta} \epsilon_1^\mu \epsilon_2^\nu q_1^\alpha q_2^\beta \, {\cal F}_{\eta_Q} (t_1,t_2) \, ,
\end{equation}
where $\alpha_{\rm em}$ is the electromagnetic coupling, $\epsilon_1$ and $\epsilon_2$ represent the polarisation vectors of the two photons, and $\varepsilon_{\mu \nu \alpha \beta}$ is the complete anti-symmetric Levi-Civita tensor. In the equation above, we have introduced the dimensionless variables
\begin{equation}
    t_1 = \frac{q_1^2}{m_Q^2},  \, \qquad \qquad  t_2 = \frac{q_2^2}{m_Q^2} \, ,
\end{equation}
where $m_Q$ is the mass of the heavy quark. These variables vanish for on-shell photons, and are negative, $t_i<0$, in the space-like, and positive, $t_i>0$, in the time-like region respectively.

In this section, we discuss the relativistic corrections in $v^2$ to the transition form-factor. We follow closely the procedure outlined in refs.~\cite{Bodwin:2002cfe,Sang:2009jc}, while keeping the full dependence on both photon virtualities. The NRQCD factorisation formula \cite{Bodwin:1994jh} for the form-factor reads
\begin{equation}
    {\cal F}_{\eta_Q}(t_1,t_2) = \frac{C_0(t_1,t_2)}{m_Q^2} \Bigbra{\eta_Q} \Big| \psi^{\dag} \chi \Big| \Bigket{0} + \frac{D_0(t_1,t_2)}{m_Q^4} \Big \langle \eta_Q \Big| \psi^{\dag} \Big(-i \frac{\tensor{\mathbf D}}{2} \Big)^2 \chi \Big| 0 \Big \rangle + \dots \, . 
    \label{eq:TFF_NRQCD}
\end{equation}
Here $C_0$, $D_0$ are the short-distance coefficients, which depend on the normalised photon virtualities $t_1,t_2$, and we explicitly show the relevant Long-Distance Matrix Elements (LDMEs) to order $v^2$. 
We can now calculate the short-distance coefficients $C_0,D_0$ by evaluating the perturbative QCD amplitude for the process $\gamma^*(q_1) \gamma^*(q_2) \to  (Q \bar Q)[^1 S^{[1]}_0]$, where the heavy quarks are in the colour-singlet and spin-singlet $S$-wave state in correspondence with the quantum numbers of the $\eta_Q$ meson.

The tree-level amplitude for the two diagrams with $t$- and $u$-channel quark exchange for photon-photon fusion reads
\begin{multline}    
{\cal M}(\gamma^*(q_1) \gamma^*(q_2) \to Q_{\lambda}(p_Q) \bar Q_{\bar \lambda}(p_{\bar Q})) = - 4\pi \alpha_{\rm em} e_Q^2\; \delta_{ij}  \\ 
\times \epsilon_1^{\mu}\,\epsilon_2^{\nu}\,\bar u_\lambda (p_Q) 
\Big( \gamma_\mu \frac{\slashed{p}_Q  - \slashed{q}_1 + m_Q}{ (p_Q - q_1)^2- m_Q^2} \gamma_\nu + 
\gamma_\nu  \frac{\slashed{p}_Q - \slashed{q}_2 + m_Q}{  (p_Q-q_2)^2 - m_Q^2} \gamma_\mu
\Big) v_{\bar \lambda}(p_{\bar Q}) \, ,
\end{multline}
where $e_Q$ is the fractional charge of the heavy quark $Q$ and the subscripts in $\delta_{ij}$ correspond to the colour indices of the heavy quarks. We write the quark and anti-quark momenta as 
\begin{equation}
    p_Q = \frac{1}{2}P +k\, , \quad    p_{\bar Q} = \frac{1}{2}P -k\, .
    \label{eq:momentumheavyquark}
\end{equation}
In the centre-of-mass frame, they take the form
\begin{equation}
     P = (2E, \vec{0})\, , \quad k = (0,\vec k) \, , \quad E = \sqrt{m_Q^2 + \vert \vec k \vert^2}\, .
     \label{eq:momentumadditional}
\end{equation}
Choosing a coordinate system, in which the $z$--axis points along the $\gamma^* \gamma^*$ collision axis, the four-momenta of the photons read
\begin{equation}
    q_1 = (\omega_1, \vec{0}_{\perp},q_z)\, , \quad q_2 = (\omega_2, \vec{0}_{\perp},  -q_z)\, .
\end{equation}
The photon energies are given by
\begin{equation}
    \omega_1 = \frac{M_{\eta_Q}^2 + q_1^2 - q_2^2}{2 M_{\eta_Q}}\, , \quad \omega_2 = \frac{M_{\eta_Q}^2 + q_2^2 - q_1^2}{2 M_{\eta_Q}}\, , \quad \omega_1 + \omega_2 = 2E \equiv M_{\eta_Q} \, ,
    \label{eq:equalityEM}
\end{equation}
where $M_{\eta_Q}$ is the mass of the bound state. The $q_z$ momentum then reads
\begin{equation}
    q_z = \frac{\sqrt{X}}{M_{\eta_Q}}, \quad  X \equiv (q_1 \cdot q_2)^2 -q^2_1 q^2_2 = \frac{1}{4}(M_{\eta_Q}^4 - 2M_{\eta_Q}^2(q^2_1 + q^2_2) + (q^2_1 - q^2_2)^2).
\end{equation}

In order to project out the $Q\bar{Q}$ pair onto the pseudo-scalar $^1S_0$ state, we amputate the spinors and apply the spin-singlet projection operator $\Pi_0$ and the colour-singlet operator $C_{ij}$  \cite{Sang:2009jc} where
\begin{equation}
\begin{split}
\Pi_0 =& \frac{1}{4 \sqrt{2} E (E + m_Q)}\left (\frac{\slashed{P}}{2} - \slashed{k} - m_Q \right) \gamma_5 \left( \slashed{P} + 2E \right) \left (\frac{\slashed{P}}{2} + \slashed{k} + m_Q \right),
\\
C_{ij} =& \frac{\delta^{ij}}{\sqrt{N_c}},
\end{split}
\end{equation}
to obtain
\begin{align}
    {\cal M} =& -4\pi \alpha_{em} e_Q^2 \, \epsilon_1^{\mu}\,\epsilon_2^{\nu}\,
    {\rm Tr}\Bigg[ \Pi_0\,   \Big( \gamma_\mu \frac{(\slashed{p}_Q  - \slashed{q}_1 + m_Q)}{q_1^2 - 2 p_Q \cdot q_1} \gamma_\nu + 
\gamma_\nu  \frac{(\slashed{p}_Q - \slashed{q}_2 + m_Q)}{ q_2^2 - 2 p_Q \cdot q_2} \gamma_\mu
\Big) \Bigg] \frac{1}{\sqrt{N_c}} {\rm Tr}[\mathbb{I}_c]
\nonumber \\
        =& -i\, 4 \pi \alpha_{em} e^2_Q \sqrt{N_c} \frac{\sqrt{2}m_Q}{E} \Big(   \frac{1}{D_1} +  \frac{1}{D_2} \Big) \, \varepsilon_{\mu\nu \alpha\beta}\, \epsilon^\mu_1 \epsilon^\nu_2 q^\alpha_1 q^\beta_2\, .
\end{align}
Above, $N_c=3$ is the number of colours. The denominators $D_1$ and $D_2$ depend on the polar angle $\theta$ of the quark relative momentum $\vec k$ with respect to the photon-photon collision axis,
\begin{equation}
\begin{split}
    D_1 =& q_1^2 - 2 p_Q \cdot q_1= - \Big( 2E^2 - \frac{1}{2} (q_1^2 + q_2^2)  - \frac{|\vec k| \sqrt{X}}{E} \cos \theta\Big) \,, 
    \\
   D_2 =& q_2^2 - 2 p_Q \cdot q_2 = - \Big( 2E^2 - \frac{1}{2} (q_1^2 + q_2^2)  + \frac{|\vec k| \sqrt{X}}{E} \cos \theta\Big) .
\end{split}
\end{equation}
In order to project on the $S$-wave, we average over the angles of $\vec k$,
\begin{equation}
    \tilde{A_0} \equiv \int \frac{d \Omega(\vec k)}{4 \pi} \Big( \frac{1}{D_1} + \frac{1}{D_2} \Big) = 
    \frac{1}{2} \int^{1}_{-1} d\cos \theta \,  \Big( \frac{1}{D_1} + \frac{1}{D_2} \Big) = -\frac{E}{|\vec k| \sqrt{X}} {\rm ln} \Big( \frac{a_+}{a_-}  \Big)\, ,
    \label{eq:s_wave_average}
\end{equation}
with
\begin{equation}
    a_\pm =  2E^2 - \frac{1}{2} (q_1^2 + q_2^2)  \pm  \frac{|\vec k| \sqrt{X}}{E}.
\end{equation}
Our amplitude then takes the form
\begin{equation}
    {\cal M} =  4 \pi \alpha_{em} \, i \, \varepsilon_{\mu \nu \alpha \beta}\, \epsilon_1^\mu \epsilon_2^\nu q_1^\alpha q_2^\beta  \,  {\cal  F}_{^1S_0}(t_1,t_2) \, , 
    \end{equation}
with the form-factor:
\begin{equation}
 {\cal F}_{^1S_0}(t_1,t_2) =  -e_Q^2 \sqrt{2N_c} \, \frac{m_Q}{E} \tilde{A_0}  = 
e^2_Q \sqrt{2N_c} \frac{m_Q}{|\vec{k}|\sqrt{X}} {\rm ln} \Big( \frac{a_+}{a_-} \Big) \, ,
\label{eq:FF_sd_all_order}
\end{equation}
which includes the full dependence on $\vert \vec k \vert / m_Q$.

In order to expand the transition form-factor into a series in $\vert \vec k \vert^2/m_Q^2$, we find it convenient to introduce the combinations of $t_1, t_2$,
\begin{equation}
 \tau = - \frac{t_1 + t_2}{4} \, , \quad \omega = \frac{t_1- t_2}{t_1 + t_2} \, . 
\label{eq:tauomegadef}
\end{equation}
While $\omega \in [-1,1]$, the symmetric variable $\tau$ is positive in the space-like region and vanishes at the double on-shell point.
We can now perform an expansion in powers of $\vert \vec k \vert^2/m_Q^2$ as
\begin{multline}
{\cal F}_{^1S_0}(t_1,t_2) = e_Q^2 \sqrt{2N_c} \frac{1}{m_Q^2} 
    \frac{1}{1+ \tau}
    \Bigg[ 
    1 + \frac{\vert \vec k \vert^2}{m_Q^2} \, \frac{2 \omega^2\tau^2 -2\tau -3(\tau+1)^2 -4}{6(\tau + 1)^2}
    \\
    + \bigg(\frac{\vert \vec k \vert^2}{m_Q^2}\bigg)^2 \, \frac{1}{(\tau+1)^4} \Bigg(  \frac{3}{8}(\tau +1)^4  + \frac{1}{2}(\tau +1)^3 +\frac{1}{6}(\tau +1)^2(-3\omega^2\tau^2 - 2\tau +7)
    \\
    -(\tau+1)(\omega^2\tau^2 +2\tau +1) + \frac{1}{5}(\omega^2\tau^2 +2\tau +1)^2 \Bigg) 
    + {\cal O} \bigg(  \bigg(\frac{\vert \vec k \vert^2}{m_Q^2}\bigg)^3  \bigg)
    \Bigg] \, . 
    \label{eq:FF_pQCD_expanded}
\end{multline}
We observe that, while, at the lowest order, the form-factor is only a function of $\tau$, the dependence on $\omega$ appears already at the first order $\vert \vec k \vert^2/m_Q^2$ in the relativistic expansion.

We now turn to the matching to the NRQCD formula in eq.~(\ref{eq:TFF_NRQCD}) and the extraction of the short-distance coefficients $C_0$, $D_0$ at order $\vert \vec k \vert^2/m_Q^2$.
We therefore replace the hadronic state in eq.~(\ref{eq:TFF_NRQCD}) by the perturbative colour-singlet, spin-singlet $S$-wave state of the $Q \bar Q$ pair and write
\begin{equation}
\begin{split}
    {\cal F}_{^1S_0}(t_1,t_2) =& \frac{C_0(t_1,t_2)}{m_Q^2} \Bigbra{(Q \bar Q)[^1S^{[1]}_0]} \Big| \psi^{\dag} \chi \Big| \Bigket{0}
    \\
    & + \frac{D_0(t_1,t_2)}{m_Q^4} \Big \langle (Q \bar Q)[^1S^{[1]}_0] \Big| \psi^{\dag} \Big(-i \frac{\tensor{{\mathbf D}}}{2} \Big)^2 \chi \Big| 0 \Big \rangle + \dots \, .
    \label{eq:FF_pQCD}
\end{split}
\end{equation}
The perturbative matrix elements in the expansion of eq.~(\ref{eq:FF_pQCD}) are easily evaluated to be~\cite{Sang:2009jc},
\begin{align}
    \Bigbra{(Q \bar Q)[^1S^{[1]}_0]} \Big| \psi^{\dag} \chi  \Big| \Bigket{0 } =& \sqrt{2N_c} \, 2 E\, ,
    \\
    \Bigbra{(Q\bar Q)[^1S^{[1]}_0]} \Big| \psi^{\dag} \Big(-i \frac{\tensor{\mathbf D}}{2} \Big)^2 \chi \Big| \Bigket{0 } =& \sqrt{2N_c}\, 2E \, \vert \vec{k} \vert^2 \, , 
\end{align}
where the factor $\sqrt{2N_c}$ stems from spin and colour factors in the normalisation of the $(Q \bar Q)[^1S^{[1]}_0]$ state, while the factor $2E$ stems from the relativistic normalisation of states adopted by us.
Inserting these results into eq.~(\ref{eq:FF_pQCD}) and expanding in $\vert \vec k \vert^2/m_Q^2$, we obtain
\begin{equation}
   {\cal F}_{^1S_0}(t_1,t_2) = 
   \frac{2 \sqrt{2N_c}}{m_Q} \Big\{ C_0(t_1,t_2)  + \frac{\vert \vec k \vert^2}{m_Q^2} \Big( \frac{1}{2} C_0(t_1,t_2)+ D_0(t_1,t_2) \Big) + \dots\Big\} \, . 
\end{equation}
Comparing with eq.~(\ref{eq:FF_pQCD_expanded}), we identify the short-distance coefficients as
\begin{equation}
    C_0(t_1,t_2) = \frac{e_Q^2}{2 m_Q} \, \frac{1}{1+\tau} \, , \quad  D_0(t_1,t_2) = \frac{\omega^2 \tau^2 - 3 \tau^2 - 7 \tau -5}{3 (1+\tau)^2} \, C_0(t_1,t_2) \, .
\end{equation}
Let us define the effective expansion parameter
\begin{equation}
    \langle v^2 \rangle_{\eta_Q} = \frac{\Bigbra{ \eta_Q} \Big| \psi^{\dag} \Big(-i \frac{\tensor{\mathbf D}}{2} \Big)^2 \chi \Big| \Bigket{0} }{m_Q^2 \Bigbra{\eta_Q} \Big| \psi^{\dag} \chi \Big| \Bigket{0} } = \frac{\langle \vert \vec k \vert^2 \rangle_{\eta_Q}}{m_Q^2} \, ,
\label{eq:relksqvsq}
\end{equation}
so that we obtain for the transition form-factor at the hadron level in eq.~(\ref{eq:TFF_NRQCD}),
\begin{equation}
\begin{split}
     {\cal F}_{\eta_Q}(t_1,t_2) =& \frac{1}{m_Q^2}  \, \Bigbra{\eta_Q}  \Big| \psi^{\dag} \chi \Big| \Bigket{0} \, 
     \Big( C_0(t_1,t_2) + D_0(t_1,t_2) \langle v^2 \rangle_{\eta_Q} + \dots \Big)
     \\
     =& {\cal F}_0(t_1,t_2)\,  \Big( 1 + \frac{\omega^2 \tau^2 - 3 \tau^2 - 7 \tau -5}{3 (1+\tau)^2} \, \langle v^2 \rangle_{\eta_Q} + \dots \Big) 
     \, ,
\end{split}
\end{equation}
where we introduced the Leading-Order (LO) form-factor
\begin{equation}
    {\cal F}_0(t_1,t_2) = \frac{e_Q^2}{2m_Q^3}  \frac{1}{1+\tau} \Bigbra{\eta_Q} \Big| \psi^{\dag} \chi \Big| \Bigket{0} \, . 
    \label{eq:TFF_LO}
\end{equation}

In a similar fashion, it is possible to perform a matching to operators with higher derivative orders. In Appendix~\ref{sec:appendixresummation}, we show how this can be done consistently at higher orders and how the relativistic corrections can be resummed to all orders in $\langle v^2 \rangle_{\eta_Q}$. Here we quote the final result
\begin{equation}
\begin{split}
      {\cal F}_{\eta_Q}(t_1,t_2) =& \frac{  \bra{\eta_Q} | \psi^{\dag} \chi | \ket{0} }{2 \sqrt{2N_c}} \, \frac{ {\cal F}_{^1S_0}(t_1,t_2,\langle v^2 \rangle_{\eta_Q})}{E}
      \\
      =& {\cal F}_0(t_1,t_2) \left(1 + g(t_1,t_2, \langle v^2 \rangle_{\eta_Q}) \right) ,
\end{split}
\end{equation}
where we have defined
\begin{equation}
 g(t_1,t_2, \langle v^2 \rangle_{\eta_Q}) = \, \frac{m_Q^2}{\sqrt{X} }\frac{(1+\tau)}{\sqrt{\langle v^2 \rangle_{\eta_Q} + \langle v^2 \rangle_{\eta_Q}^2}} {\rm ln} \Big( \frac{a_+}{a_-} \Big) - 1  \, .
 \label{eq:g_t1t2}
\end{equation}
It is understood that there is an implicit dependence on $\langle v^2 \rangle_{\eta_Q}$ in $X$ and $a_{\pm}$, where $\vert \vec{k} \vert^2$ has to be replaced by $\langle \vert \vec{k} \vert^2 \rangle_{\eta_Q}$. In addition, we note that, after expressing $X$ and $a_{\pm}$ in terms of $t_1$, $t_2$ and $\langle v^2 \rangle_{\eta_Q}$, the function $g$ has no explicit dependence on $m_Q$.

\begin{figure}
  \begin{center}
    \subfloat[]{\includegraphics[width=6.5cm]{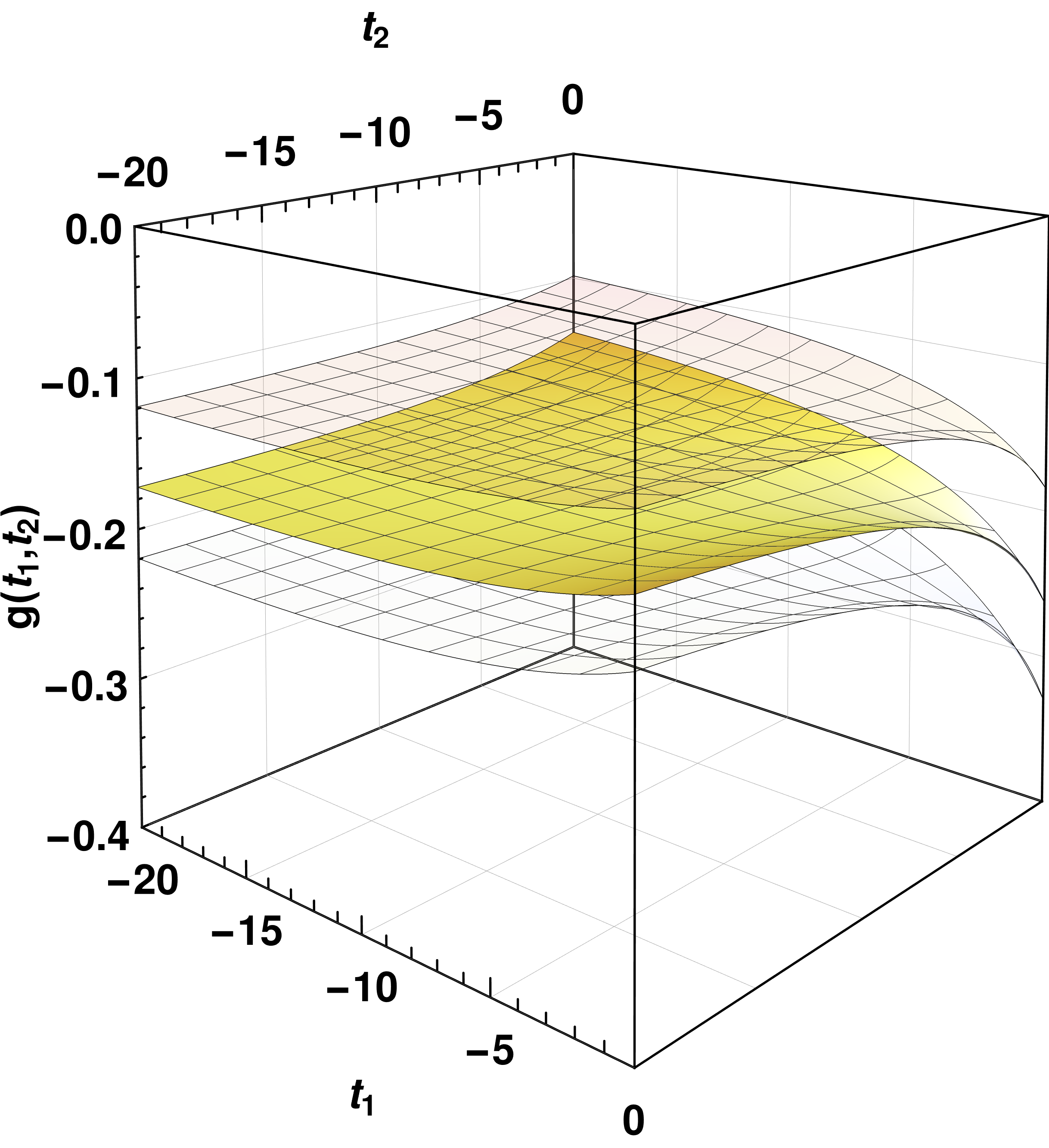} \label{fig:g_t1t2}}
    \quad\quad
    \subfloat[]{\includegraphics[width=6.5cm]{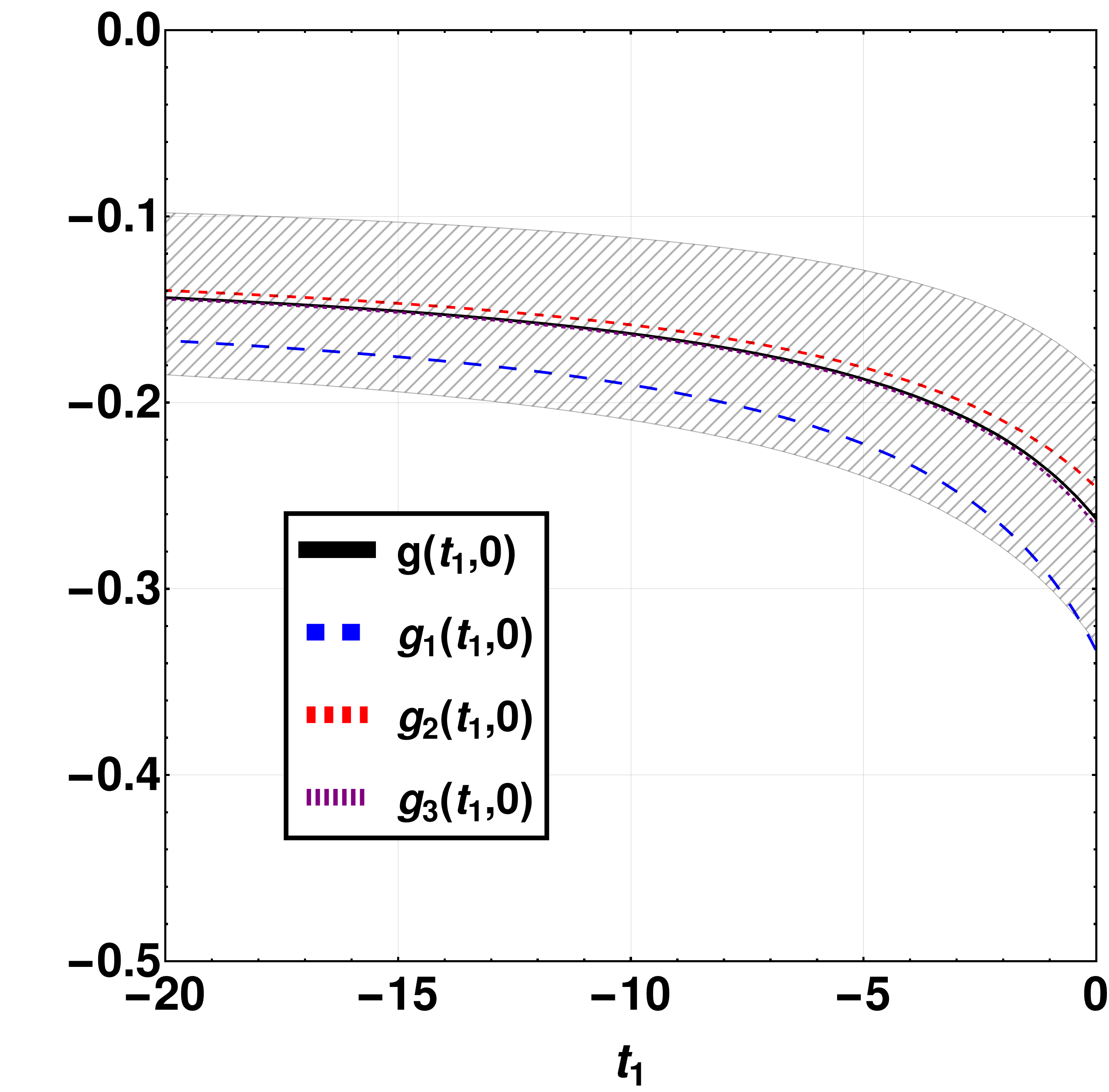} \label{fig:g_t1impact}}
  \end{center}
  \caption{(a) Plot of $g(t_1, t_2)$ (see eq.~(\ref{eq:g_t1t2})) as a function of $t_1$ and $t_2$ and (b) plot of $g(t_1, 0)$ and $g_n(t_1, 0)$ (see eq.~(\ref{eq:g_n_t1t2})) with $n=\{1,2,3\}$ as a function of $t_1$ only. We set $\langle v^2 \rangle_{\eta_c} = 0.20 \pm 0.07$ as default parameter for the charmonium case, that will be discussed later in Section~\ref{Sec:Applications}, and show the uncertainties of $g$ only.}
  \label{fig:g_t1t2both}
\end{figure}

In Fig.~\ref{fig:g_t1t2}, we show $g(t_1,t_2, \langle v^2 \rangle_{\eta_Q})$ as a function of $t_1$ and $t_2$ with parameters as described in the figure caption for the case of $\eta_c$. We see that relativistic corrections in $g$ are largely negative over the phase-space considered and can reach corrections of the order of 10\% to 30\% compared to the leading order term.

In order to understand the impact of the first few terms in the $\langle v^2 \rangle_{\eta_Q}$ expansion, we define
\begin{equation}
 g_n(t_1,t_2, \langle v^2 \rangle_{\eta_Q}) = \, \sum_{q=1}^{n} \frac{\langle v^2 \rangle_{\eta_Q}^q}{q!} \left.\left(\frac{\partial^q}{\partial \langle v^2 \rangle^q} \, g(t_1,t_2, \langle v^2 \rangle) \right)\right\vert_{\langle v^2 \rangle = 0} \, .
 \label{eq:g_n_t1t2}
\end{equation}
In Fig.~\ref{fig:g_t1impact}, we display the function $g_n(t_1,0, \langle v^2 \rangle_{\eta_Q})$ with $n=\{1,2,3\}$ and the full $g(t_1,0, \langle v^2 \rangle_{\eta_Q})$ as a function of $t_1$ only. We observe that the corrections oscillate and rapidly converge to the full $g(t_1,0, \langle v^2 \rangle_{\eta_Q})$ after the third order.

We note that, in the derivation of eq.~(\ref{eq:g_t1t2}) (see Appendix~\ref{sec:appendixresummation}) and in line with ref.~\cite{Bodwin:2007fz}, we have made use of the generalised Gremm-Kapustin relation derived in ref.~\cite{Bodwin:2006dn} which implies that
\begin{equation}
   \langle \vert\vec k\vert^{2n} \rangle_{\eta_Q} = \langle \vert\vec k\vert^2  \rangle^n_{\eta_Q} \, .
\label{eq:implicGK}
\end{equation}
In principle, there exist corrections to the generalised Gremm-Kapustin relation which are, however, suppressed by relative corrections in $v^2$. We believe the impact of such additional corrections on $g(t_1, t_2, \langle v^2 \rangle_{\eta_Q})$ to be rather small, starting at the third order in $\langle v^2 \rangle_{\eta_Q}$, which is already covered by the uncertainties of $\langle v^2 \rangle_{\eta_Q}$ (see Fig.~\ref{fig:g_t1impact}).

Before concluding this section, we point out that the operator matrix element can be expressed in terms of the radial wavefunction at the origin $R_{\eta_Q}{(0)}$. The operator matrix element uses a relativistic normalisation of states. It can be converted to the non-relativistic Bodwin-Braaten-Lepage (BBL) convention \cite{Bodwin:1994jh,Braaten:1998au} via
\begin{equation}
\begin{split}
   \bra{\eta_Q}|\psi^\dagger \chi |\ket{0}  =& \sqrt{2M_{\eta_Q}}\bra{\eta_Q}|\psi^\dagger \chi |\ket{0}_{\rm BBL} = 
 \sqrt{2M_{\eta_Q}} \sqrt{\frac{N_c}{2 \pi}}  R_{\eta_Q}{(0)}
   \\
=& \left(1 + \langle v^2 \rangle_{\eta_Q}\right)^{\frac{1}{4}} \sqrt{\frac{2 N_c m_Q}{\pi}}  R_{\eta_Q}{(0)}\, ,
\end{split}
\label{eq:LDMER0}
\end{equation}
where, in the second line, we made use of eq.~(\ref{eq:momentumadditional}), eq.~(\ref{eq:equalityEM}) and
\begin{equation}
    \vert \vec{k} \vert^2 = \langle \vert \vec{k} \vert^2 \rangle_{\eta_Q}
\end{equation}
to express $M_{\eta_Q}$ in terms of $m_Q$ and $\langle v^2 \rangle_{\eta_Q}$.

This concludes the discussion of the relativistic corrections in the NRQCD expansion, and we now turn to the discussion of the perturbative corrections to the transition form-factor.

\section{NNLO QCD corrections to $\gamma^* \gamma^* \leftrightarrow {^1S_0}$ transition form-factor}
\label{Sec:NNLO}

In this section, we discuss the Next-to-Next-to-Leading Order (NNLO) perturbative QCD corrections to the transition form-factor $\gamma^*{\left(q_1\right)} \gamma^*{\left(q_2\right)} \leftrightarrow {^1S_0}{\left(P\right)}$. We can express the perturbative correction to the form-factor as follows
\begin{equation}
    \mathcal{F}_{\eta_Q}{\left(t_1,t_2\right)}=\mathcal{F}_0{\left(t_1,t_2\right)}\left(1 + \Delta{(t_1, t_2)}\right) \, ,
\end{equation}
where $\mathcal{F}_0$ is the LO form-factor and $\Delta$ is the QCD correction factor. It admits an expansion in the strong coupling constant $\alpha_s$ as follows
\begin{equation}
    \Delta{(t_1, t_2)} = \left( \frac{\alpha_s}{\pi} \right) K_1{\left(t_1,t_2\right)} + \left(\frac{\alpha_s}{\pi}\right)^2 K_2{\left(t_1,t_2\right)} + \mathcal{O}{\left(\alpha_s^3\right)} \, ,
\label{eq:deltalabelp}
\end{equation}
where $K_1$ is the Next-to-Leading Order (NLO) correction factor, while $K_2$ is the NNLO contribution.

\begin{figure}
  \begin{center}
    \subfloat[]{\includegraphics[width=4.2cm]{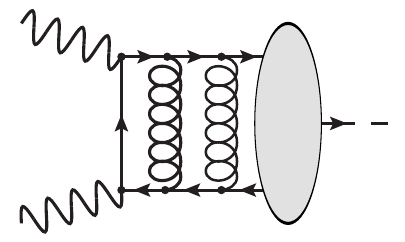}
    \label{fig:classreg}}\, 
    \subfloat[]{\includegraphics[width=4.9cm]{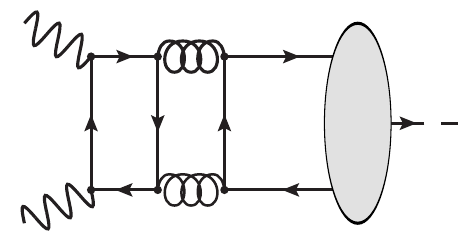} 
    \label{fig:classlbl}}\, 
    \subfloat[]{\includegraphics[width=4.5cm]{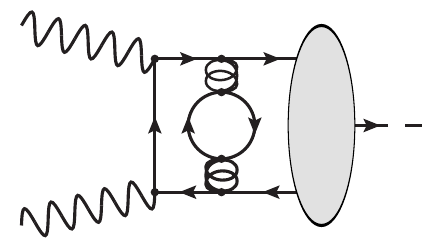} 
    \label{fig:classvac}}
  \end{center}
  \caption{Two-loop diagrams for the form-factor 
  $\gamma^* \gamma^* \leftrightarrow {^1S_0}$ with 
  (a) regular contributions, (b) light-by-light contributions and 
  (c) vacuum polarisation contributions.}
  \label{fig:twoloopdiagrams}
\end{figure}

In order to compute the amplitudes at NLO and NNLO for this exclusive process, we only need to take into account virtual loop corrections. In Fig.~\ref{fig:twoloopdiagrams} we show some representative two-loop diagrams that contribute at NNLO. We can organise and classify these diagrams into three different sets of gauge-invariant contributions, the regular two-loop contributions as in Fig.~\ref{fig:classreg}, which are free from any fermion loops, the light-by-light contributions as depicted in Fig.~\ref{fig:classlbl}, where the fermion loop is connected to the initial-state photons, and the vacuum polarisation contribution as in Fig.~\ref{fig:classvac}, where the fermion loop is attached to the gluon propagator. 

We use the package $\mathtt{FeynArts}$ to generate the Feynman diagrams and corresponding amplitudes and the package $\mathtt{FeynCalc}$ to deal with their evaluation \cite{Hahn:2000kx, Shtabovenko:2016sxi, Shtabovenko:2020gxv, Shtabovenko:2023idz}. In order to regulate singularities, we employ dimensional regularisation with $D=4-2\epsilon$.

For the QCD corrections, we work at leading order in the $v^2$-expansion, and hence take the corresponding kinematic limits in eq.~(\ref{eq:momentumheavyquark}) and eq.~(\ref{eq:momentumadditional}), $k=0$ and $E=m_Q$, which lead to degenerate heavy-quark momenta
\begin{equation}
    p_Q=p_{\bar Q} = \frac{1}{2}\left(q_1 + q_2\right) \, .
\end{equation}
As such, there are only two independent momenta, which we choose to be $q_1$ and $q_2$.

In these degenerate kinematics, some Feynman integrals exhibit linearly-dependent propagators. This linear dependence can be eliminated via the well-known procedure of partial-fraction decomposition. In order to facilitate this step, we use the package $\mathtt{Apart}$ \cite{Feng:2012iq}. For details and implications of partial-fraction decomposition, we guide the reader to Section~2.2.1 in ref.~\cite{Abreu:2022vei}. 

Having assigned all Feynman integrals to topologies with sets of linearly independent propagators, we can then use Integration-By-Parts (IBP) identities to reduce the Feynman integrals to the so-called master integrals \cite{Chetyrkin:1979bj,Chetyrkin:1981qh}. We apply the packages $\mathtt{FIRE}$ \cite{Smirnov:2008iw,Smirnov:2023yhb}, $\mathtt{LiteRed}$ \cite{Lee:2012cn,Lee:2013mka} and $\mathtt{Kira}$ \cite{Klappert:2020nbg} for the IBP reduction.

At one-loop level, we find that the amplitude can be expressed in terms of 7 master integrals of which we need only 4 of them, as the others can be deduced from these via symmetry relations (e.g. $t_1 \leftrightarrow t_2$). The computed analytical results of these one-loop integrals via direct integration have been cross-checked numerically via $\mathtt{pySecDec}$ \cite{Borowka:2017idc} in both time-like and space-like regions for $t_1$ and $t_2$.

At two-loop level, we establish that the amplitude with two off-shell photons can be expressed in terms of 199 master integrals. In the special case with only one off-shell photon, the number of master integrals reduces to 136. In the double on-shell case, as considered in ref.~\cite{Abreu:2022vei,Abreu:2022cco}, the amplitude was expressible in terms of 61 master integrals.

For this double on-shell case, one of us had computed in ref.~\cite{Abreu:2022vei} these master integrals analytically in terms of special functions as the multiple polylogarithms (MPLs) \cite{Goncharov:1998kja,Goncharov:2001iea}, and their elliptic generalisation, the elliptic multiple polylogarithms (eMPLs) \cite{brown2013multiple, Broedel:2014vla, Broedel:2017kkb}, and their special subset of iterated integrals of Eisenstein series \cite{Brown:mmv,ManinModular}.

While the properties of MPLs are quite well understood in the literature and their numerical evaluation straightforward with tools such as $\mathtt{GiNaC}$ \cite{Vollinga:2004sn}, the situation is quite different for the elliptic counterpart. In the special subset of iterated integrals of Eisenstein series, there exist algorithms that provide reliable and stable numerical evaluation \cite{Duhr:2019rrs}. However, for the generic case of eMPLs, their numerical evaluation turns out to be quite tedious and cumbersome with the existing methods.  

Returning to the off-shell case, the master integrals depend on two independent parameters, $t_1$ and $t_2$, hence we anticipate more involved special functions with multiple elliptic curves. As the numerical evaluation of analytic eMPL functions is not straightforward, in this work, we proceed to compute these master integrals only numerically using the auxiliary mass flow approach \cite{Liu:2017jxz, Liu:2021wks} implemented in the package $\mathtt{AMFlow}$ \cite{Liu:2022chg}. We note also alternative numerical tools such as $\mathtt{DiffExp}$ \cite{Hidding:2020ytt}, $\mathtt{SeaSyde}$ \cite{Armadillo:2022ugh} and $\mathtt{LINE}$ \cite{Prisco:2025wqs}.

We computed all master integrals to the required order in $\epsilon$ and to a precision of at least $30$ digits for a large range of our phase-space region relevant for the phenomenological applications considered in this work. After plugging in the master integrals, we note that the gauge-invariant subset of light-by-light contributions is already finite and free from any divergences. In contrast to this, as expected, the bare amplitudes for the regular and vacuum polarisation contributions contain UV singularities.

While the heavy-quark wavefunction and the heavy-quark mass are renormalised in the on-shell scheme, we employ the $\overline{\text{MS}}$-scheme for the strong coupling constant \cite{Barnreuther:2013qvf,Broadhurst:1993mw,Melnikov:2000qh,Mitov:2006xs,Bernreuther:1981sg}. For details and steps regarding the renormalisation procedure, we guide the reader to Section~3 in ref.~\cite{Abreu:2022cco}.\footnote{Note also Section~4 in ref.~\cite{Ozcelik:2025uln} for a slightly more general discussion on UV renormalisation.}

After incorporating the UV counter-terms, we find that the two-loop amplitude still contains a simple pole in $\epsilon$, which we identify with the known Coulomb singularity in the literature \cite{Czarnecki:2001zc,Kniehl:2006qw,Feng:2015uha}. We remove this singularity through an equivalent renormalisation of the LDME in the $\overline{\text{MS}}$-scheme \cite{Abreu:2022cco}.

We are now in a position to present the results for the finite and divergence-free amplitudes. Before doing so, we first define the colour factors that appear in the results,
\begin{equation}
    C_F = \frac{N_c^2-1}{2 N_c} \, , \qquad C_A = N_c\, , \qquad T_F = \frac{1}{2}\, ,
\end{equation}
where $N_c=3$ is the number of colours. In addition, we indicate with $n_l$ the number of light quark flavours running in the coupling constant, and with $n_h$ the number of massive quarks that appear in the fermion loops.

At one-loop level, the correction factor can be expressed as
\begin{equation}
    K_1{(t_1,t_2)} = C_F \, a_F{(t_1, t_2)}\, ,
    \label{eq:K1oneloopresult}
\end{equation}
while at two-loop level the structure is more involved and reads
\begin{equation}
\begin{split}
    K_2{(t_1,t_2)} = & K_{2;\textrm{reg}}{(t_1,t_2)} + K_{2;\textrm{lbl}}{(t_1,t_2)} + K_{2;\textrm{vac}}{(t_1,t_2)}
    \\
    & + \frac{1}{4}\, \beta_0 \, K_1{(t_1,t_2)}\, l_{\mu_R} + \frac{1}{2}\, \gamma_{\textrm{Coulomb}}\, l_{\mu_{\Lambda}} \, ,
\end{split}
\label{eq:K2twoloopresult}
\end{equation}
where we have introduced the short-hand notation
\begin{equation}
    l_{\mu} = \log{\left(\frac{\mu^2}{m_Q^2}\right)}
\end{equation}
to indicate the scale-dependence for the renormalisation scale $\mu_R$ and the NRQCD scale $\mu_{\Lambda}$. The associated anomalous-dimension coefficients are
\begin{align}
    \beta_0 =& \frac{11}{3}\, C_A - \frac{4}{3}\, T_F\, n_l \, ,
    \\
    \gamma_{\textrm{Coulomb}} =& -\pi^2 \left(C_F^2 + \frac{1}{2}\, C_F C_A \right) \, .
\end{align}

In eq.~(\ref{eq:K2twoloopresult}), the quantities $K_{2;\textrm{reg}}$, $K_{2;\textrm{lbl}}$ and $K_{2;\textrm{vac}}$ represent the regular, light-by-light and vacuum polarisation contributions respectively. These can be structured further as follows
\begin{align}
    K_{2;\textrm{reg}}{(t_1,t_2)} =& C_F^2 \, a_{FF}{(t_1, t_2)} + C_F C_A \, a_{FA}{(t_1, t_2)} \, ,
    \\[5pt]
    K_{2;\textrm{lbl}}{(t_1,t_2)} =& C_F T_F n_h \, b_{Fh}{(t_1,t_2)} + C_F T_F \tilde{n}_l \, b_{Fl}{(t_1,t_2)} \, ,
    \\[5pt]
    K_{2;\textrm{vac}}{(t_1,t_2)} =& C_F T_F n_h \, c_{Fh}{(t_1,t_2)} + C_F T_F n_l \, c_{Fl}{(t_1,t_2)} \, ,
\end{align}
where above we have introduced
\begin{equation}
    \tilde{n}_l = \sum_i^{n_l} \frac{e_i^2}{e_Q^2}
\end{equation}
to take into account the difference in the fractional electric charge for the different quark flavours.

We have performed a number of cross-checks on the two-loop amplitude. We were able to verify that all poles cancel after UV renormalisation and the Coulomb singularity subtraction. Further, we had computed the amplitude at several kinematic points that mapped to other points under exchange of $t_1$ and $t_2$. We are hence able to confirm that the amplitude is indeed symmetric under exchange of $t_1$ and $t_2$.

In ref.~\cite{Feng:2015uha}, the authors have computed the transition form-factor with a single off-shell photon and provided some benchmark results for the two-loop coefficient at different values of $t_1$. They provided results for the combined regular and vacuum polarisation contributions and separately also for the light-by-light contributions. We have cross-checked their results against ours and we find agreement. Furthermore we have checked that we are able to reproduce the known result of the double on-shell case computed in ref.~\cite{Abreu:2022cco,Czarnecki:2001zc} in the limit $t_{1} \rightarrow 0$, $t_{2} \rightarrow 0$.

\renewcommand{\arraystretch}{1.5}
\begin{table}[h]
    \centering
    \begin{tabular}{| l || c | c | c |}
    \hline
                 & $t_1=-\frac{3}{5}$, $t_2=-\frac{3}{4}$ & $t_1=-\frac{23}{5}$, $t_2=0$ & $t_1=-\frac{69}{10}$, $t_2=-\frac{211}{30}$ \\
      \hline
      \hline
      $a_{FF}$   & $-20.67715$ & $-20.17373$ & $-19.09761$ \\
      \hline
      $a_{FA}$   & $-4.86457$ & $-4.64160$ & $-5.78452$ \\
      \hline
      $b_{Fh}$   & $0.65669 + i\, 2.16026$ & $0.64507 + i\, 2.29462$ & $0.40649 + i\, 2.30711$ \\
      \hline
      $b_{Fl}$   & $-0.26215 - i\, 1.99684$ & $-0.84202 - i\, 2.25641$ & $-1.58848 - i\, 1.70059$ \\
      \hline
      $c_{Fh}$   & $0.21271$ & $0.19687$ & $0.10679$ \\
      \hline
      $c_{Fl}$   & $-0.57472$ & $-0.71757$ & $-0.34842$ \\
      \hline
    \end{tabular}
    \caption{Benchmark results for the different coefficients at three selected kinematic points.}
    \label{tab:benchmark}
\end{table}

In Table~\ref{tab:benchmark}, we provide some benchmark results for the different two-loop coefficients accurate up to 5 digits after the decimal for three different kinematic points. In Appendix~\ref{sec:appendixanalyticalK1}, we give the analytical results for the $K_1$ coefficient that is valid for both space-like and time-like regions.

\begin{figure}
    \begin{center}
    \subfloat[]{\includegraphics[width=5.2cm]{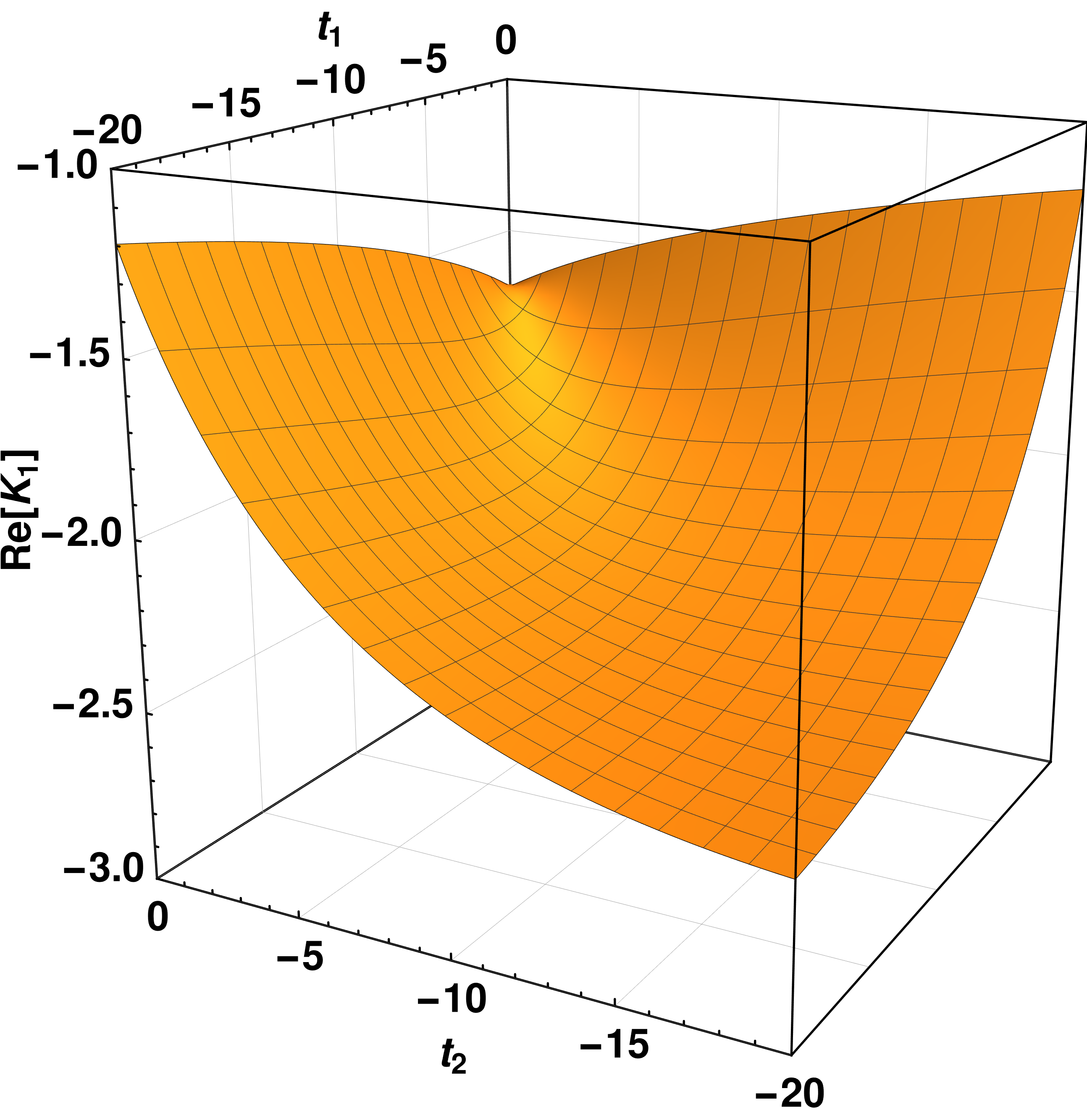}}\quad
    \subfloat[]{\includegraphics[width=5.2cm]{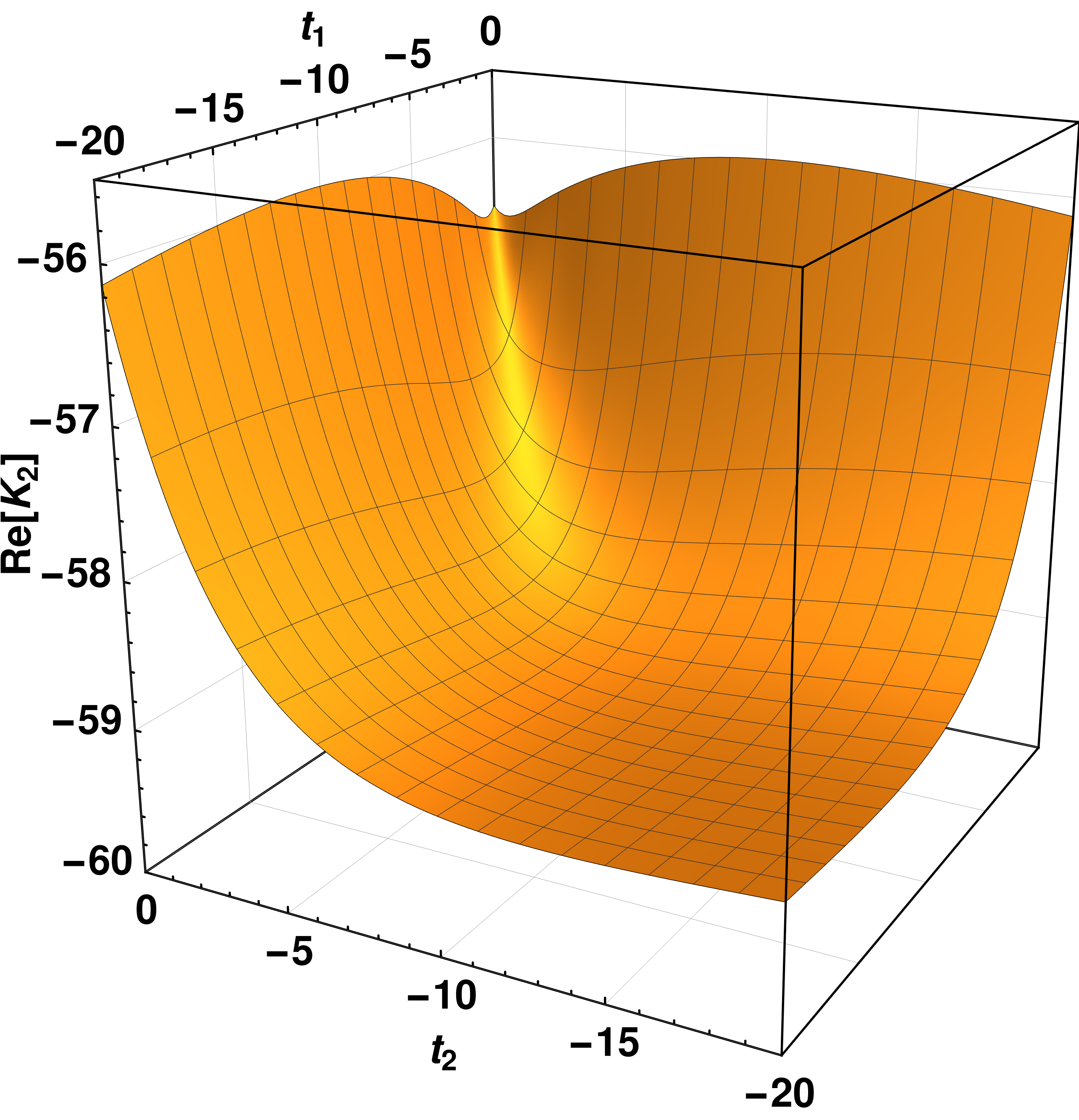}}\quad
    \subfloat[]{\includegraphics[width=5.2cm]{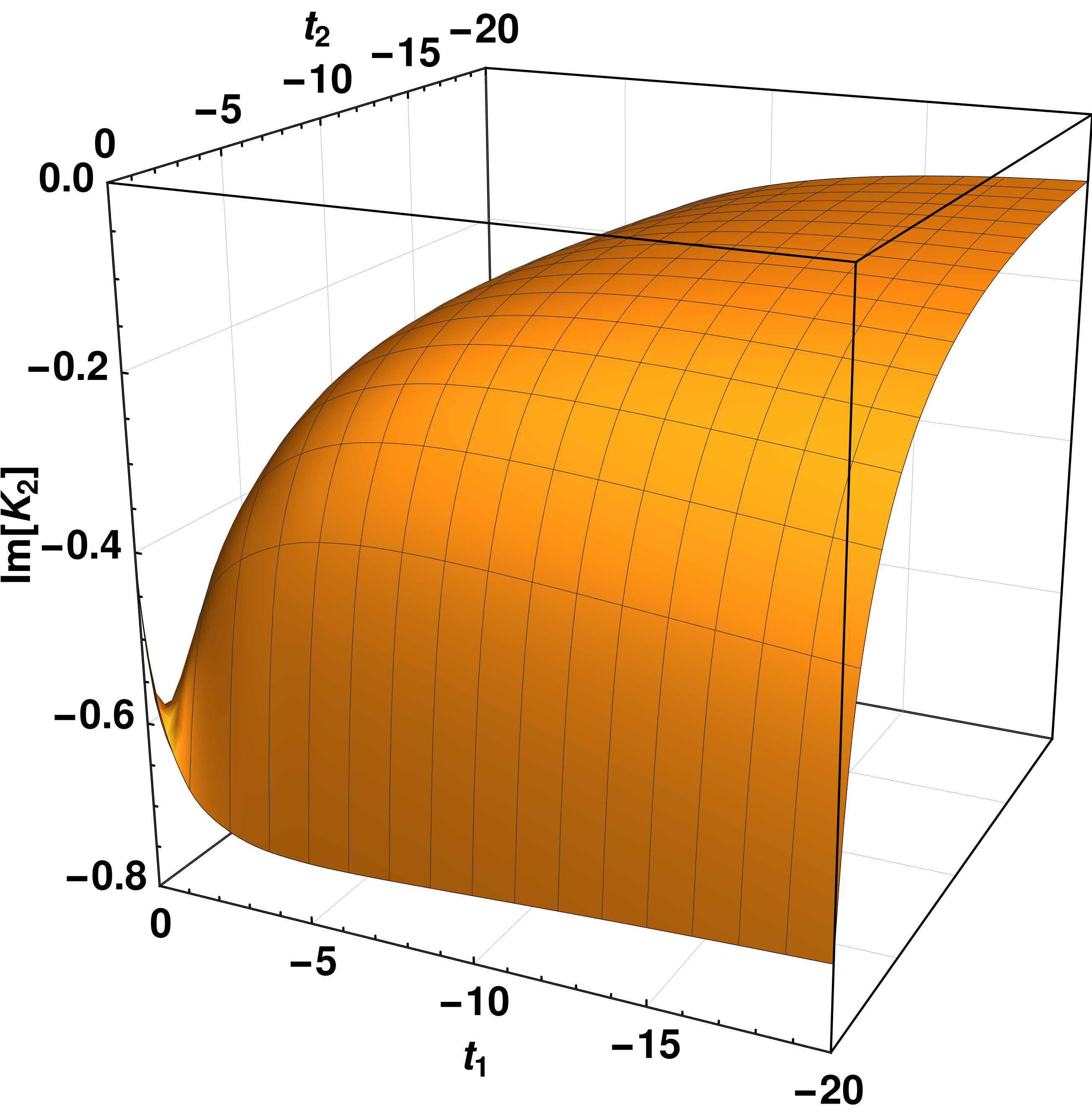}}
    \end{center}
    \caption{NLO and NNLO correction factors as a function of $t_1$ and $t_2$ in the double space-like region. We plot in (a) the real part of $K_{1}(t_1, t_2)$, while in (b) and (c) we plot the real and imaginary parts of $K_{2}(t_1, t_2)$ respectively. For $K_2(t_1, t_2)$, we set $l_{\mu_R}=0$, $l_{\mu_{\Lambda}}=0$, $n_h=1$, $n_l=3$ and $\tilde{n}_l=3/2$ relevant for the charmonium case.}
    \label{fig:K1K2_factor}
\end{figure}

In Fig.~\ref{fig:K1K2_factor}, we display the shape of $K_1{(t_1, t_2)}$ and $K_2{(t_1, t_2)}$ as a function of $t_1$ and $t_2$ for two space-like photons. For the NNLO correction factor we set in the plot $l_{\mu_R}=0$, $l_{\mu_{\Lambda}}=0$ and further set the number of flavours to $n_h=1$, $n_l=3$ and $\tilde{n}_l=3/2$ relevant for the $\eta_c$ case.\footnote{For the bottomonium $\eta_b$ case, one would have instead $n_l=4$ and $\tilde{n}_l=10$, where the light-by-light contribution part is significantly enhanced compared to the $\eta_c$ case. In the toponium $\eta_t$ case, which has received recent attention at the LHC \cite{CMS:2025kzt,ATLAS:2025mvr}, the values read $n_l=5$ and $\tilde{n}_l=11/4$. \label{foot:bottomflavnum}} While the $K_1$ function is purely real, the $K_2$ function exhibits both a real and an imaginary component. The origin of the imaginary contribution can be traced back to the light-by-light contribution.

We observe that both the NLO coefficient function $\Ree{\left[K_1\right]}$ and the NNLO coefficient function $\Ree{\left[K_2\right]}$ are negative in the phase-space considered. While the double on-shell point starts, in the case of $\Ree{\left[K_1\right]}$, at $\mathcal{O}{\left(-1\right)}$, it is much more negative for $\Ree{\left[K_2\right]}$ at $\mathcal{O}{\left(-50\right)}$. We also note that the variation across $t_1$ and $t_2$ is stronger in the NNLO case. As for $\Ime{\left[K_2\right]}$, it is negative as well in the same phase-space region with the double on-shell point at $\mathcal{O}{\left(-0.5\right)}$ and has large gradients close to the on-shell points.

\begin{figure}
  \begin{center}
    \subfloat[]{\includegraphics[width=6.5cm]{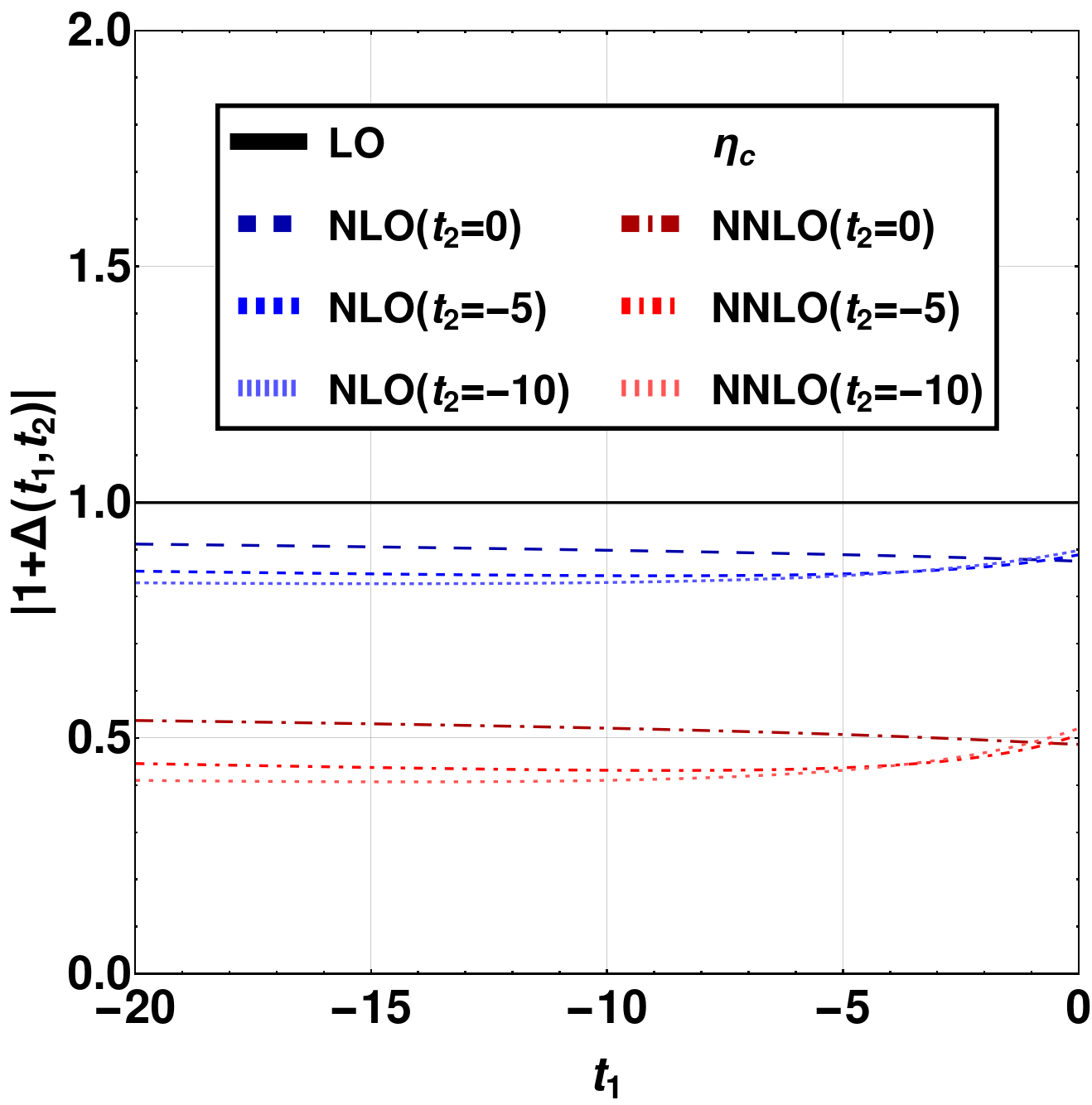} \label{fig:pQCDetac}}
    \quad\quad\quad
    \subfloat[]{\includegraphics[width=6.5cm]{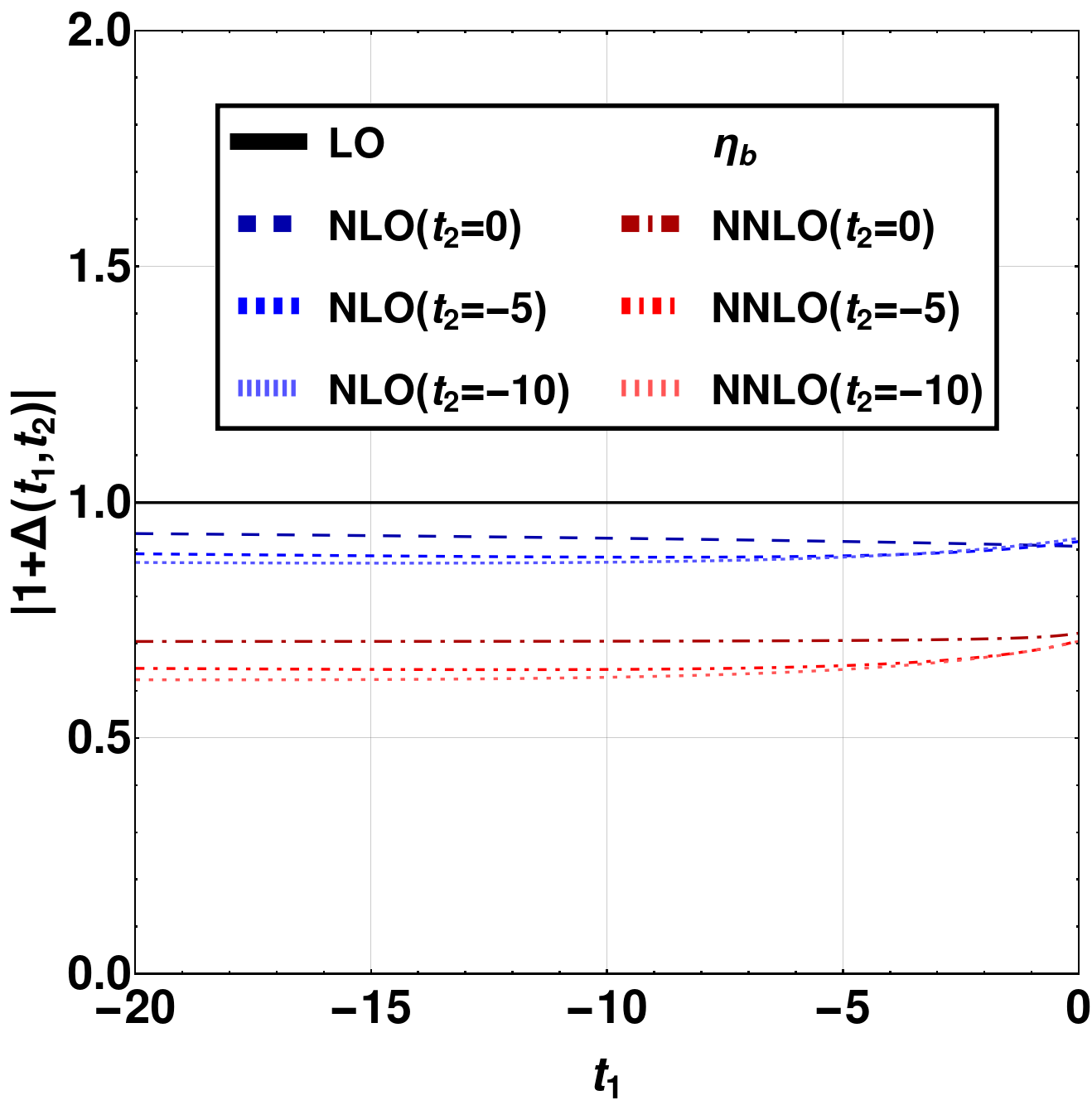} \label{fig:pQCDetab}}
  \end{center}
  \caption{Plots of $\left\vert 1 + \Delta(t_1, t_2) \right\vert$ (see eq.~(\ref{eq:deltalabelp})) at LO, NLO and NNLO for (a) $\eta_c$ and (b) $\eta_b$ as a function of $t_1$ and at three different values of $t_2=\{0,-5,-10\}$. We set $l_{\mu_R}=\log{4}$ with $\mu_R^2 = 4m_Q^2$, and $l_{\mu_{\Lambda}}=0$. While we use $m_c=1.5 \text{ GeV}$ for the $\eta_c$ case, we set $m_b=4.8 \text{ GeV}$ for $\eta_b$ and adjust the number of light flavours (see footnote~\ref{foot:bottomflavnum}).}
  \label{fig:pQCDetacandb}
\end{figure}

In Fig.~\ref{fig:pQCDetacandb}, we illustrate the size of the QCD corrections in $\left\vert 1 + \Delta(t_1, t_2) \right\vert$ for both the $\eta_c$ and the $\eta_b$ cases as a function of $t_1$ and at three different values of $t_2$ with parameters as described in the figure caption. While the LO curve is by construction unity at any kinematic point, NLO corrections are negative in the ranges considered and can reach corrections of the order of up to $20\%$ for $\eta_c$ and up to $15\%$ for $\eta_b$. We note that the NNLO corrections are large and are negative as well and can yield, in addition to the NLO terms, an additional correction of up to $40\%$ and $25\%$ respectively for the $\eta_c$ and $\eta_b$ cases. Lowering the $\mu_{\Lambda}$ scale would decrease the size of the NNLO term.

We conclude this section by noting that, in addition to the $K_2$ factor, also the LDME $\bra{\eta_Q} | \psi^{\dag} \chi | \ket{0}$ exhibits a dependence on the NRQCD scale $\mu_{\Lambda}$. The dependence can be derived from the fact that at each order in $\alpha_s$, the transition form-factor $\mathcal{F}_{\eta_Q}{\left(t_1,t_2\right)}$ has to be free from the scale. The evolution equation reads\footnote{Note that, in the case of inclusion of mixed QCD and relativistic corrections, there are additional terms in the evolution at $\mathcal{O}{\left(\alpha_s \langle v^2 \rangle\right)}$ \cite{Bodwin:1994jh,Guo:2011tz,Jia:2011ah}. \label{fn:evolutionQCDrelativistic}}
\begin{equation}
    \frac{\partial \bra{\eta_Q} | \psi^{\dag} \chi | \ket{0} {\left(\mu_{\Lambda}\right)}}{\partial \log{\mu^2_{\Lambda}}} = - \frac{1}{2} \left(\frac{\alpha_s{\left(\mu_{\Lambda}\right)}}{\pi}\right)^2  \gamma_{\textrm{Coulomb}} \, \bra{\eta_Q} | \psi^{\dag} \chi | \ket{0} {\left(\mu_{\Lambda}\right)} + \mathcal{O}{\left(\alpha_s^3, \alpha_s \langle v^2 \rangle\right)}\, .
\end{equation}
Solving this equation, we find that
\begin{equation}
    \frac{\bra{\eta_Q} | \psi^{\dag} \chi | \ket{0} {\left(\mu_{\Lambda}\right)}}{\bra{\eta_Q} | \psi^{\dag} \chi | \ket{0} {\left(\mu_{0}\right)} } = 1 - \frac{1}{2} \left(\frac{\alpha_s{\left(\mu_{\Lambda}\right)}}{\pi}\right)^2 \gamma_{\textrm{Coulomb}}\, \log{\left(\frac{\mu_{\Lambda}^2}{\mu_{0}^2}\right)} + \mathcal{O}{\left(\alpha_s^3, \alpha_s \langle v^2 \rangle\right)}\, ,
\end{equation}
where $\mu_0$ is the reference scale of the LDME.

This completes the discussion of the perturbative QCD corrections to the transition form-factor and we are now ready to discuss their phenomenological applications.

\section{Phenomenology}
\label{Sec:Applications}

In the following, we discuss three phenomenological applications of the transition form-factor for the charmonium case and investigate corrections in $\alpha_s$ and $\langle v^2 \rangle_{\eta_c}$. Combining these corrections from the previous sections, the form-factor can be written as
\begin{equation}
      {\cal F}_{\eta_c}(t_1,t_2) = {\cal F}_0(t_1,t_2) \Big(1 + g(t_1,t_2, \langle v^2 \rangle_{\eta_c}) + \Delta(t_1, t_2) \Big).
      \label{eq:FFcombrelqcd}
\end{equation}
We first discuss the corrections to the single space-like transition form-factor, before discussing the case of the double space-like transition form-factor. We then revisit the decay width of $\eta_c$ to di-photon and discuss the impact of the relativistic and perturbative QCD corrections.

In order to assess the theory uncertainties, we take into account the uncertainties related to the charm quark mass, $m_c$, uncertainties related to relativistic parameter, $\langle v^2 \rangle_{\eta_c}$, and uncertainties in the strong coupling constant, $\alpha_s$.

We take the charm quark mass to be roughly half the mass of the $\eta_c$ meson, hence we set
\begin{equation}
    m_c = \left( 1.5 \pm 0.1 \right) \text{ GeV} \, .
    \label{eq:masscentralerror}
\end{equation}

In order to determine the relativistic parameter, $\langle v^2 \rangle_{\eta_c}$, we refer to ref.~\cite{Bodwin:2007fz}, where the authors performed a simultaneous fit of $\langle \vert\vec k\vert^2  \rangle_{\eta_c} $ and $\bra{\eta_c} \vert \psi^{\dag} \chi \vert \ket{0}$ by using experimental data for $\Gamma[\eta_c \rightarrow \gamma \gamma]$ and the relation between $\langle \vert\vec k\vert^2  \rangle_{\eta_c} $ and $\bra{\eta_c} \vert \psi^{\dag} \chi \vert \ket{0}$ from the potential model. They also included data for $\Gamma[J/\psi \rightarrow e^+e^-]$ assuming heavy-quark spin symmetry of the LDMEs, allowing a combined fit of $\langle \vert\vec k\vert^2  \rangle_{\eta_c} $ and $\bra{\eta_c} \vert \psi^{\dag} \chi \vert \ket{0}$ to both sets of decay-width data.

We note that, in their analysis \cite{Bodwin:2007fz}, they used an old PDG value (2006) for the $\eta_c$ decay width $\Gamma[\eta_c \rightarrow \gamma \gamma]_{\text{old}} = \left( 7.2 \pm 2.1 \right) \text{ keV}$ \cite{ParticleDataGroup:2006fqo}, while the current PDG value (2024) sits at $\Gamma[\eta_c \rightarrow \gamma \gamma]_{\text{PDG}} = \left( 5.1 \pm 0.4 \right) \text{ keV}$ \cite{ParticleDataGroup:2024cfk}.\footnote{We also note the recent ref.~\cite{BESIII:2024rex} where the BESIII collaboration has determined $\Gamma[\eta_c \rightarrow \gamma \gamma]_{\text{BESIII}} = \left( 11.3 \pm 1.4 \right) \text{ keV}$ which significantly deviates from the PDG value by roughly a factor of two.} As the new PDG central value (2024) is nearly identical to the lower bound of the old PDG value (2006), the detailed analysis in ref.~\cite{Bodwin:2007fz} shows that updating the PDG values affects the fit only within the error bands of $\langle \vert\vec k\vert^2  \rangle_{\eta_c} $.\footnote{We thank G.T.~Bodwin, one of the authors of ref.~\cite{Bodwin:2007fz}, for pointing this out.}

Hence, in order to compute $\langle v^2 \rangle_{\eta_c}$, we make use of eq.~(53b) in ref.~\cite{Bodwin:2007fz} for $\langle \vert\vec k\vert^2  \rangle_{\eta_c} $ and the relation in eq.~(\ref{eq:relksqvsq}) to obtain
\begin{equation}
    \langle v^2 \rangle_{\eta_c} = 0.20 \pm 0.07 \, ,
\end{equation}
where we used eq.~(\ref{eq:masscentralerror}) for the charm quark mass. For the wavefunction at the origin, using eq.~(53a) in ref.~\cite{Bodwin:2007fz}, we find
\begin{equation}
    |R_{\eta_c}{(0)}|^2 = 0.92^{+0.23}_{-0.22} \text{ GeV}^3\, .
\label{eq:R0standdef}
\end{equation}

The uncertainty in the QCD coupling, $\alpha_s$, can be estimated by varying the renormalisation scale $\mu_R$ around a central scale choice. For the scale evolution of $\alpha_s$, we make use of the $\mathtt{RunDec}$ package \cite{Herren:2017osy}. At NLO, we use the one-loop renormalisation group equation, while at NNLO, we evolve the coupling to two-loop order. Furthermore, we set the number of flavours to $n_f=n_l=3$ in the running of the coupling. In addition to the implicit $\mu_R$ dependence in the coupling, the perturbative QCD correction coefficient $\Delta$ exhibits, starting from the two-loop order, also an explicit dependence on $\mu_R$ and the NRQCD factorisation scale $\mu_{\Lambda}$.

In order to assess the theory uncertainties related to the QCD corrections, we choose the central renormalisation scale to be $\mu_{R,\rm cent.}^2 = 4m_c^2$ and vary $\mu_{R,\rm cent.}^2$ by a factor of two, similarly to what was done in ref.~\cite{Feng:2015uha}. As for the NRQCD factorisation scale dependence, we note that, in addition to the LDME and the QCD correction part, also the relativistic parameter $\langle v^2 \rangle_{\eta_c}$ can in principle have a dependence on $\mu_{\Lambda}$ through the presence of the LDMEs in eq.~(\ref{eq:relksqvsq}). However, the usage of eq.~(\ref{eq:implicGK}), which implies that $\langle v^{2n} \rangle_{\eta_c} = \langle v^2 \rangle_{\eta_c}^n$, changes the LDME power and hence complicates their scale evolution due to the absence of mixed QCD and relativistic corrections at $\mathcal{O}{(\alpha_s v^2)}$ (see footnote~\ref{fn:evolutionQCDrelativistic}). For this reason, instead of varying $\mu_{\Lambda}$ around a central value, we will perform two computations, one at $\mu_0=\mu_{\Lambda} = m_c$ and the other at $\mu_0=\mu_{\Lambda}=m_c \sqrt{\langle v^2 \rangle_{\eta_c}}$.

In order to compute the uncertainties of our observables stemming from the QCD and relativistic corrections, as these corrections are of distinct origins, and since we do not take into account correlations, we vary separately the uncertainties related to the mass $m_c$, the relativistic parameter $\langle v^2 \rangle_{\eta_c}$ and the QCD renormalisation scale $\mu_R^2$ around the central value. We establish the minimal and maximal value from this variation as the uncertainties of our observable.

\subsection{Single space-like transition form-factor}

In ref.~\cite{BaBar:2010siw}, the \textsc{Ba}\textsc{Bar} collaboration performed measurements of the process $ \gamma^* \gamma \rightarrow \eta_c$ with an off-shell initial-state photon in $e^+ e^-$ collisions, which is achieved by tagging one of the scattered leptons. From these measurements, they are able to extract the experimental result for the ratio of the $\eta_c$ transition form-factor as a function of the photon virtuality $Q_1^2$ normalised towards its on-shell point.

We define the ratio of the transition form-factors as
\begin{equation}
\begin{split}
    R{(Q_1^2)}=& \frac{\left\vert\mathcal{F}_{\eta_c}(t_1,0)\right\vert}{\left\vert\mathcal{F}_{\eta_c}(0,0)\right\vert}
    \\
    =& \frac{4}{4 - t_1} \frac{\left\vert 1 + g(t_1,0, \langle v^2 \rangle_{\eta_c}) + \Delta(t_1, 0) \right\vert}{\left\vert 1 + g(0,0, \langle v^2 \rangle_{\eta_c}) + \Delta(0, 0) \right\vert} \, ,
\end{split}
\label{eq:Rsingleobservable}
\end{equation}
where we used eq.~(\ref{eq:FFcombrelqcd}) and eq.~(\ref{eq:TFF_LO}) and expressed it as a function of the photon virtuality, $Q_1^2$, that is defined as
\begin{equation}
    Q_1^2=-m_c^2\, t_1 \, .
\end{equation}
This allows us to compare the theory prediction directly with the \textsc{Ba}\textsc{Bar} measurements. We note that this observable is independent of global factors, such as the wavefunction at the origin, $|R_{\eta_c}{(0)}|^2$. In order to compute relativistic and QCD corrections for this observable, we compute them separately both in the numerator and denominator terms of the ratio.

\begin{figure}
  \begin{center}
    \subfloat[]{\includegraphics[width=5.5cm]{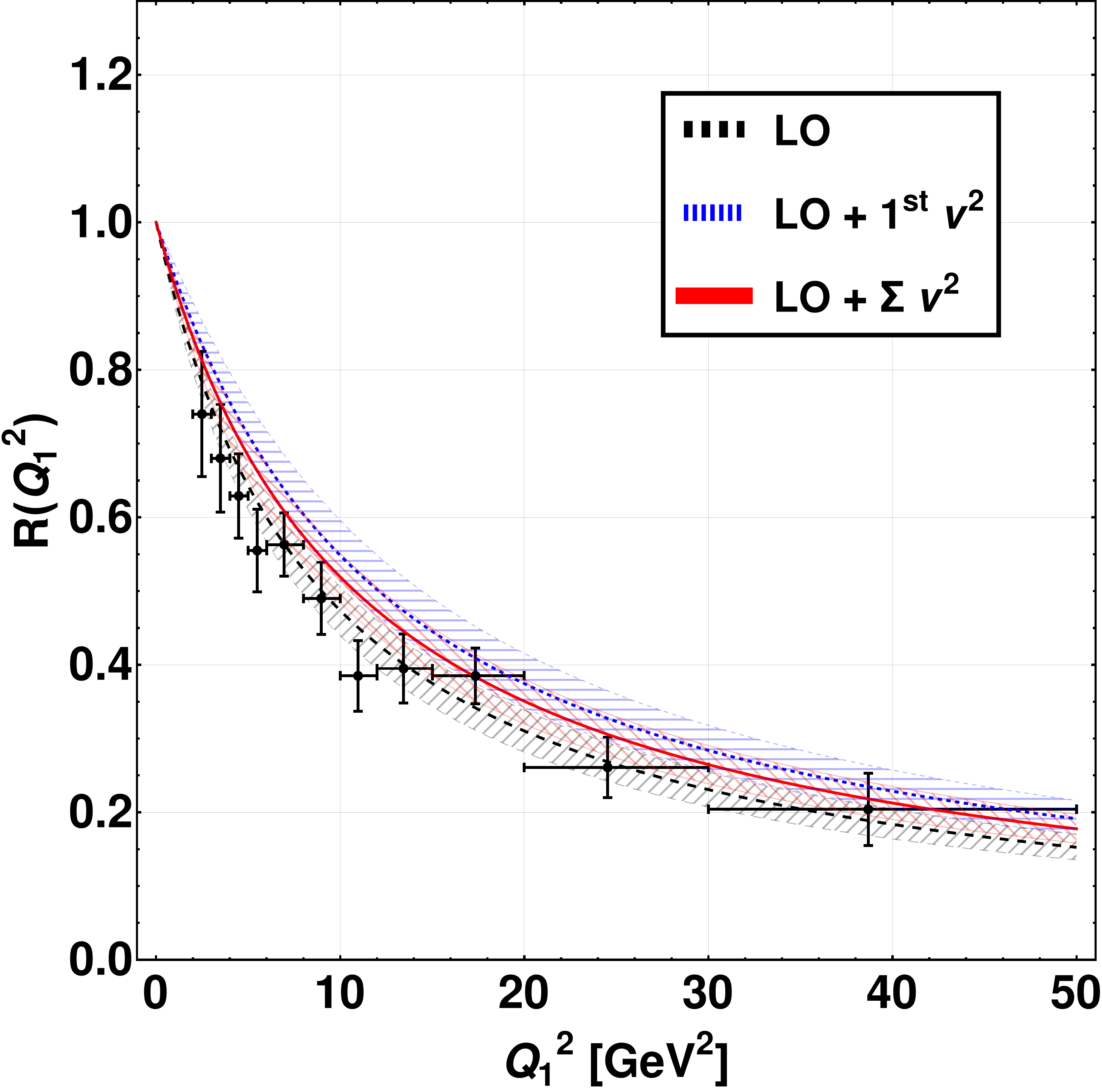} \label{fig:singleplotsrelativistic}}
    \subfloat[]{\includegraphics[width=5.5cm]{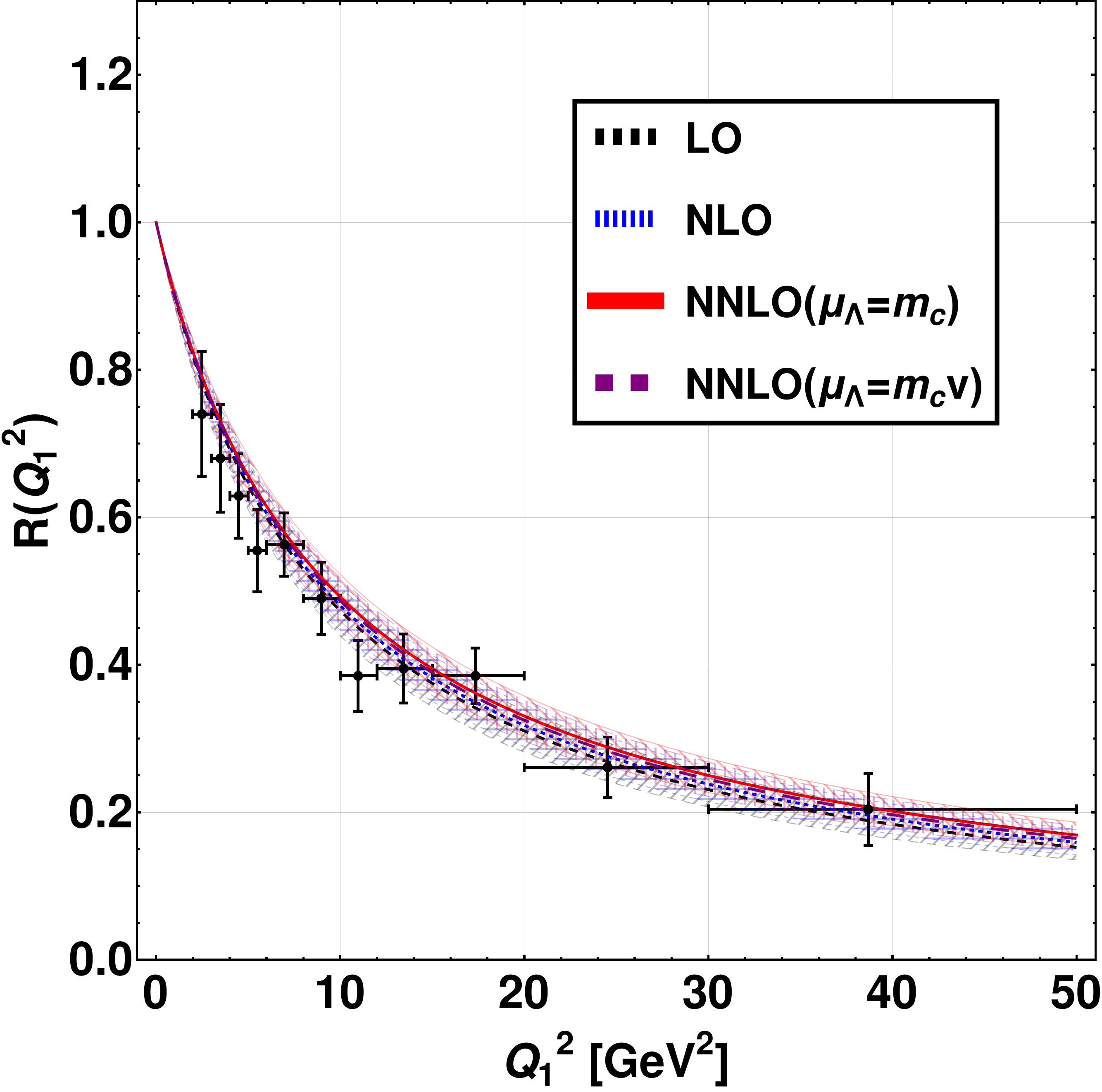} \label{fig:singleplotsQCD}}
    \subfloat[]{\includegraphics[width=5.5cm]{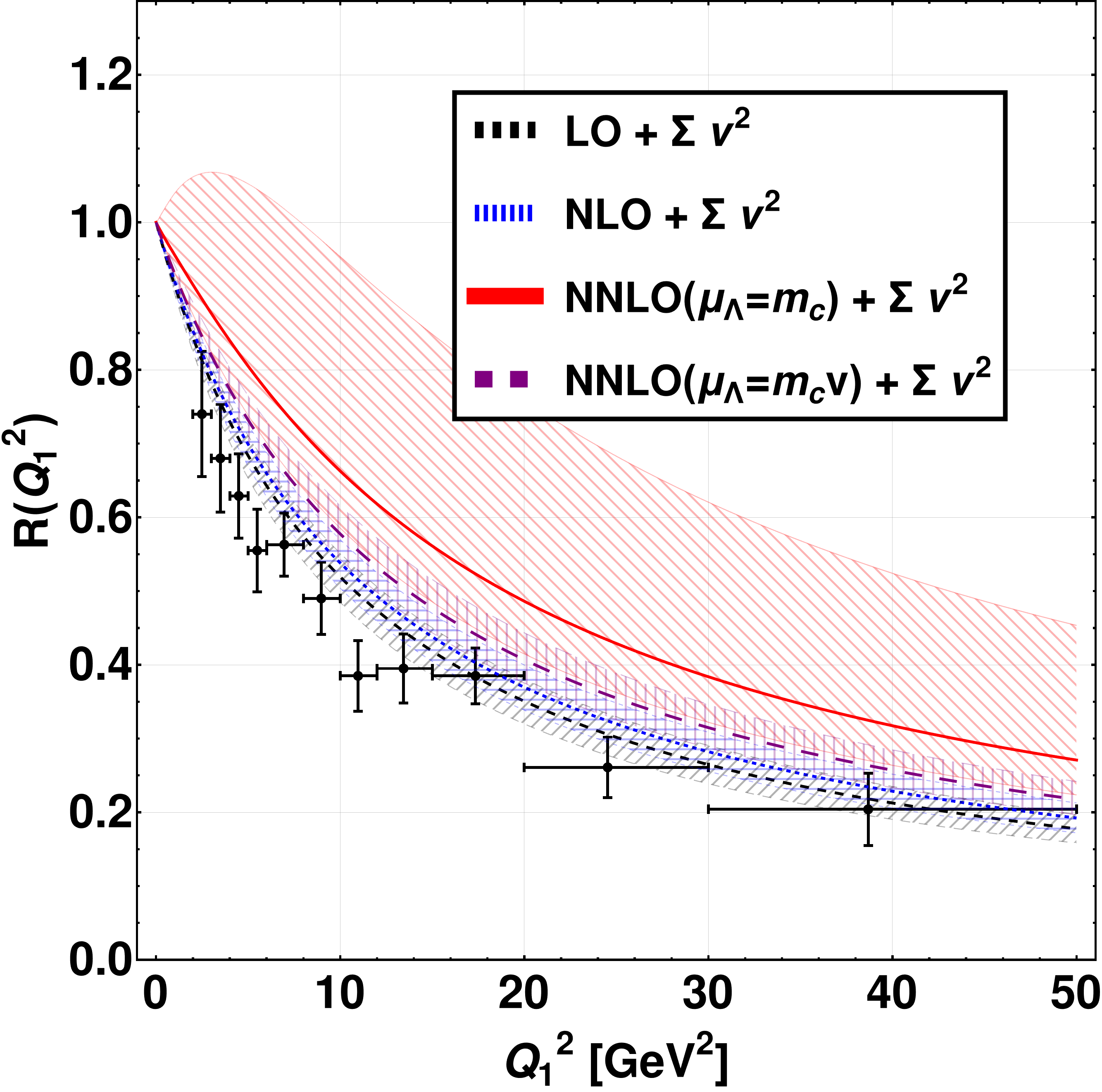} \label{fig:singleplotsQCDrelativistic}}
  \end{center}
  \caption{Transition form-factor ratio $R{(Q_1^2)}$ with impact of
  (a) relativistic corrections, (b) QCD corrections, 
  (c) combined relativistic and QCD corrections.}
  \label{fig:singleplots}
\end{figure}

In Fig.~\ref{fig:singleplots}, we show the impact of the pure relativistic corrections, the pure QCD corrections and the combined QCD and relativistic corrections on the observable $R{(Q_1^2)}$. The uncertainties have been computed according to the procedure described at the beginning of Section~\ref{Sec:Applications}. We observe that, in Fig.~\ref{fig:singleplotsrelativistic}, adding relativistic corrections pushes the LO curve up. The all-order $v^2$ resummation contribution, denoted with $\sum v^2$, reduces this effect. While the LO curve is able to describe the \textsc{Ba}\textsc{Bar} data \cite{BaBar:2010siw} quite well, the relativistic corrections, in particular the first-order relativistic correction, slightly overshoot the data.

In Fig.~\ref{fig:singleplotsQCD}, we display the effect of the pure QCD corrections at NLO and NNLO. We note that the central values of the NLO and NNLO curves push the LO curve only slightly up but are indeed within the bounds of the LO band. As such, the pure NNLO corrections turn out to be largely consistent with the \textsc{Ba}\textsc{Bar} measurements.

We notice that, in the case of the combined QCD and relativistic corrections shown in Fig.~\ref{fig:singleplotsQCDrelativistic}, all central curves are larger than the central one of the LO $+ \sum v^2$ contribution. As a consequence of this, already the combined NLO and all-order resummation contribution is slightly above the data. We also remark that the uncertainties are larger than in the case of pure QCD corrections.

In order to understand the behaviour of the combined relativistic and QCD corrections, it is instructive to look at the numerator term, $\left\vert 1 + g(t_1,0, \langle v^2 \rangle_{\eta_c}) + \Delta(t_1, 0) \right\vert$, in eq.~(\ref{eq:Rsingleobservable}). We have already shown the impact of the pure relativistic corrections, $g(t_1,0, \langle v^2 \rangle_{\eta_c})$, in Fig.~\ref{fig:g_t1impact}. While the corrections are relatively strong at the on-shell point, $t_1=0$, the corrections decrease with increasing virtuality of the photon.

\begin{figure}
  \begin{center}
    \subfloat[]{\includegraphics[width=5.5cm]{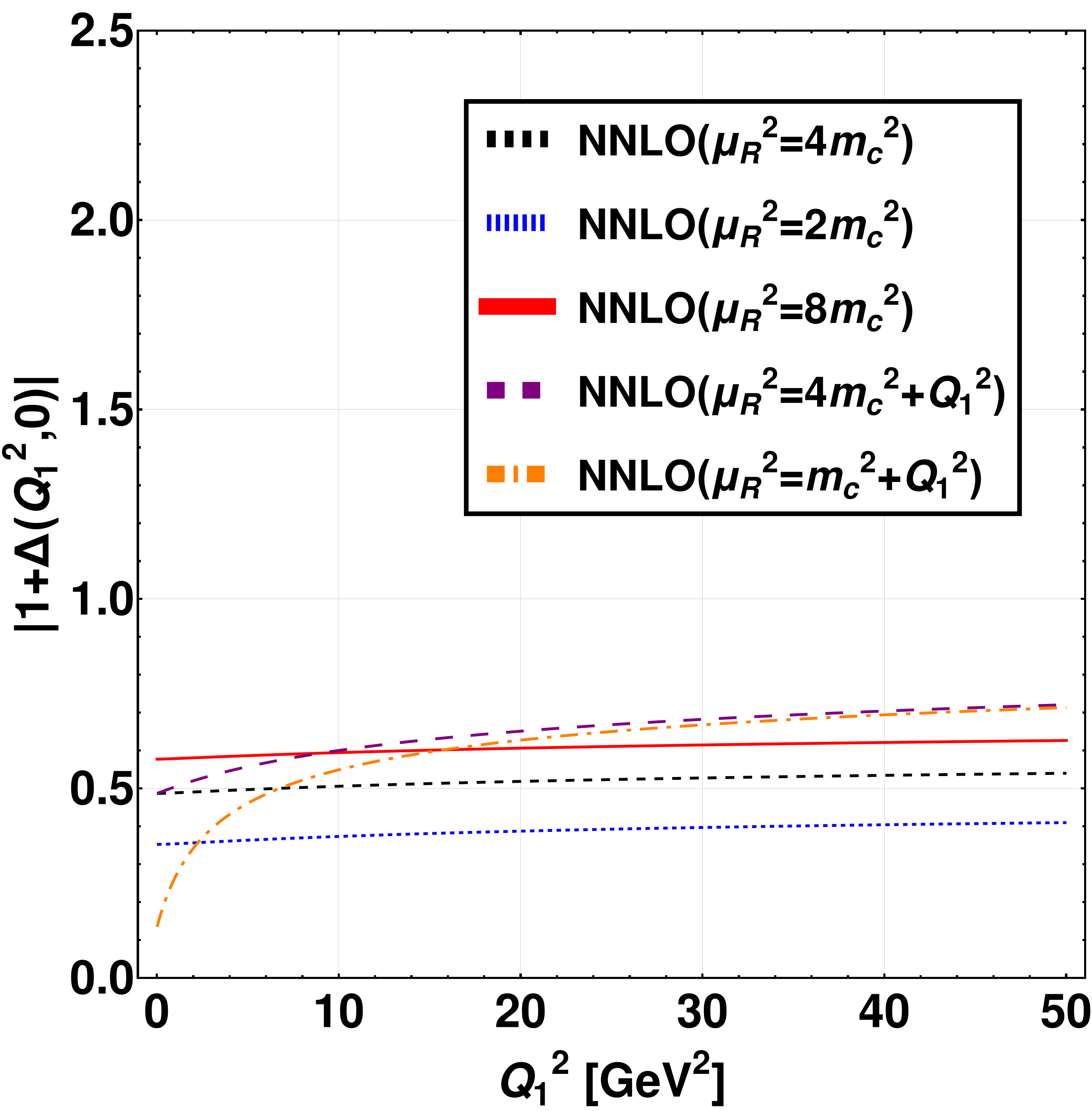} \label{fig:singlenumnnloplots}}
    \hspace{1cm}
    \subfloat[]{\includegraphics[width=5.5cm]{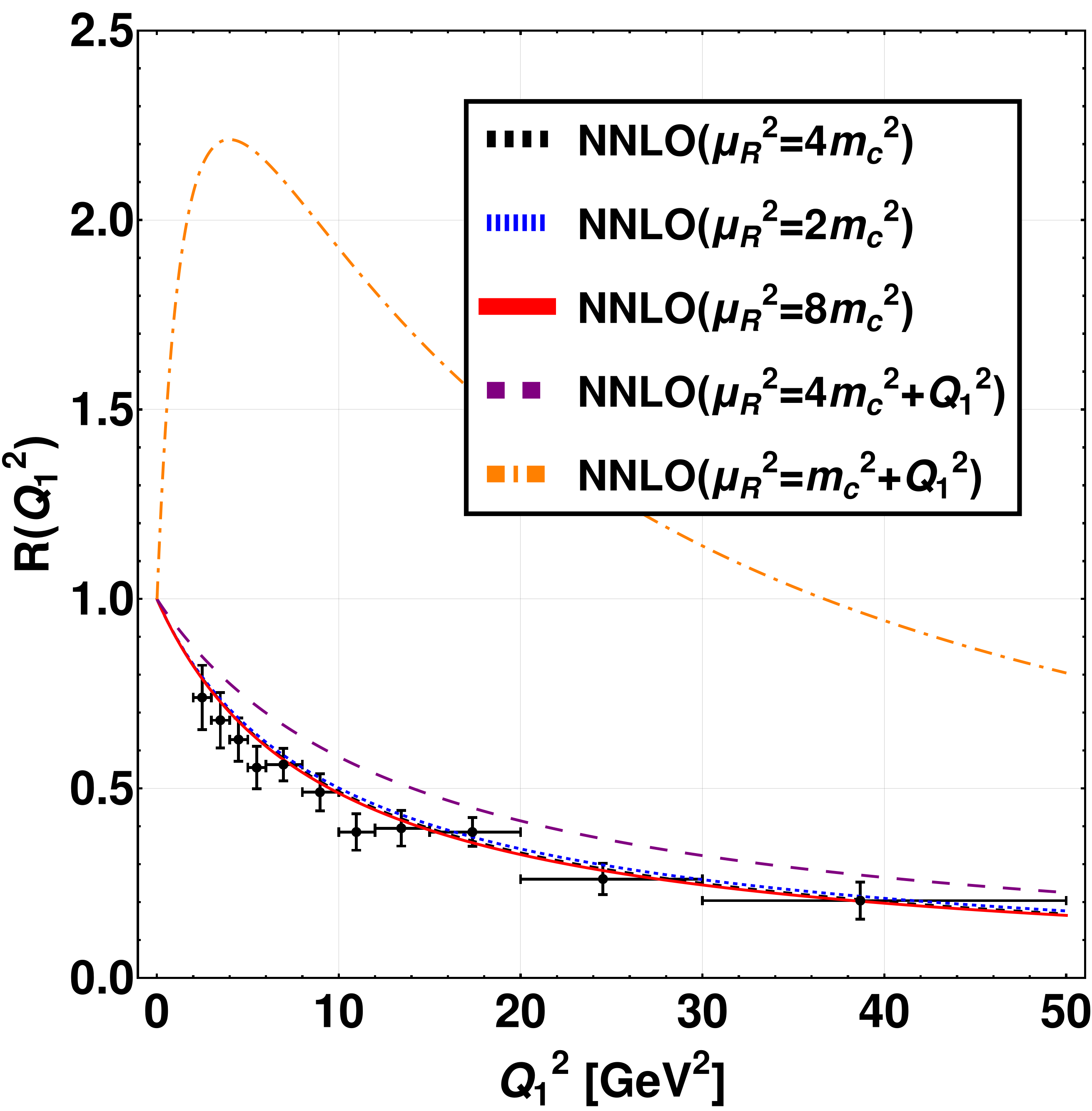} \label{fig:singleRnnloplots}}
  \end{center}
  \caption{ 
  Plots of (a) $\left\vert 1 + \Delta(Q_1^2, 0) \right\vert$ and (b) $R{(Q_1^2)}$ at NNLO with $\mu_{\Lambda}=m_c$ and different renormalisation scale choices.}
  \label{fig:singleNNLOscaleplots}
\end{figure}

In the case of the pure QCD corrections, we note that, with our renormalisation scale choice, $\mu_R$, the corrections are largely similar across different $Q_1^2$ as can be seen in Fig.~\ref{fig:singlenumnnloplots}. As a consequence of this, in the ratio to the on-shell point in eq.~(\ref{eq:Rsingleobservable}), the QCD corrections do not vary the LO curve as much as the relativistic corrections do as is visible in Fig.~\ref{fig:singleplots}.

We would like to briefly comment on ref.~\cite{Feng:2015uha}, where the authors considered the same observable, but have chosen a dynamic renormalisation scale with $\mu_R^2 = m_c^2 + Q_1^2$ where the charm quark mass was set to $m_c=1.68 \text{ GeV}$. We believe that such a renormalisation scale setting starting at the charm quark mass is rather too low, a more natural running scale might be to adopt instead $\mu_R^2 = 4 m_c^2 + Q_1^2$, which appears in the prefactor of the LO term and starts at the mass of the bound state instead.

We plot in Fig.~\ref{fig:singleNNLOscaleplots} also the results for the scale choices $\mu_R^2 = m_c^2 + Q_1^2$ and $\mu_R^2 = 4 m_c^2 + Q_1^2$ where we fix $\mu_{\Lambda} = m_c$. As is clear, due to the strong gradient close to the on-shell point, the resulting ratios are pushing the LO curves significantly up and away from the \textsc{Ba}\textsc{Bar} data (see Fig.~\ref{fig:singleRnnloplots}). Decreasing $\mu_{\Lambda}$ would weaken the QCD corrections in $\left\vert 1 + \Delta(Q_1^2, 0) \right\vert$ and consequently change the peak size in the $R{(Q_1^2)}$ ratio. We believe that this observable is best described with a fixed $\mu_R$ scale instead, in both numerator and denominator terms.\footnote{We note also ref.~\cite{Wang:2018lry, Wang:2025afy}, where the authors propose to fix $\mu_R$ by applying the Principle of Maximum Conformality (PMC) \cite{Brodsky:2011ig}. Their method results in a larger scale, $\mu_R^2 > m_c^2$, with rather flat $Q_1^2$-dependence close to the on-shell point, thereby incidentally avoiding the issues encountered in ref.~\cite{Feng:2015uha}.}

We note that, in the \textsc{Ba}\textsc{Bar} measurements \cite{BaBar:2010siw}, the form-factor at the on-shell point, $\mathcal{F}_{\eta_c}{(0,0)}$, is extracted with a slightly different method using data with non-tagged electrons, while the form-factors with non-vanishing virtualities are extracted from data with single-tagged electrons.

As described in their ref.~\cite{BaBar:2010siw}, they use different detection efficiency parameters for the non-tagged and single-tagged events. We believe that it may be worthwhile to study the following observable with a different normalisation instead
\begin{equation}
\begin{split}
    \mathcal{R}{(Q_1^2)}=& \frac{\left\vert\mathcal{F}_{\eta_c}(t_1,0)\right\vert}{\left\vert\mathcal{F}_{\eta_c}(\tilde{t}_1,0)\right\vert}
    \\
    =& \frac{4 - \tilde{t}_1}{4 - t_1} \frac{\left\vert 1 + g(t_1,0, \langle v^2 \rangle_{\eta_c}) + \Delta(t_1, 0) \right\vert}{\left\vert 1 + g(\tilde{t}_1,0, \langle v^2 \rangle_{\eta_c}) + \Delta(\tilde{t}_1, 0) \right\vert} \, ,
\end{split}
\label{eq:Rsingleobservablealternative}
\end{equation}
where
\begin{equation}
    \tilde{t}_1 = -\frac{2.49 \text{ GeV}^2}{m_c^2}
\end{equation}
represents the virtuality of the lowest non-vanishing bin. By definition, we have that $\mathcal{R}{(2.49 \text{ GeV}^2)} = 1$.

By considering this alternative observable, we have effectively eliminated the dependence on the efficiency parameter for the non-tagged events. We also remark here that there is negligible correlation among the data points, hence, in this new normalisation, the statistical uncertainty on the first bin can be safely propagated to all other bins in the error analysis.\footnote{We thank V.~Druzhinin from the \textsc{Ba}\textsc{Bar} collaboration for clarifying this point.}

\begin{figure}
  \begin{center}
    \subfloat[]{\includegraphics[width=5.5cm]{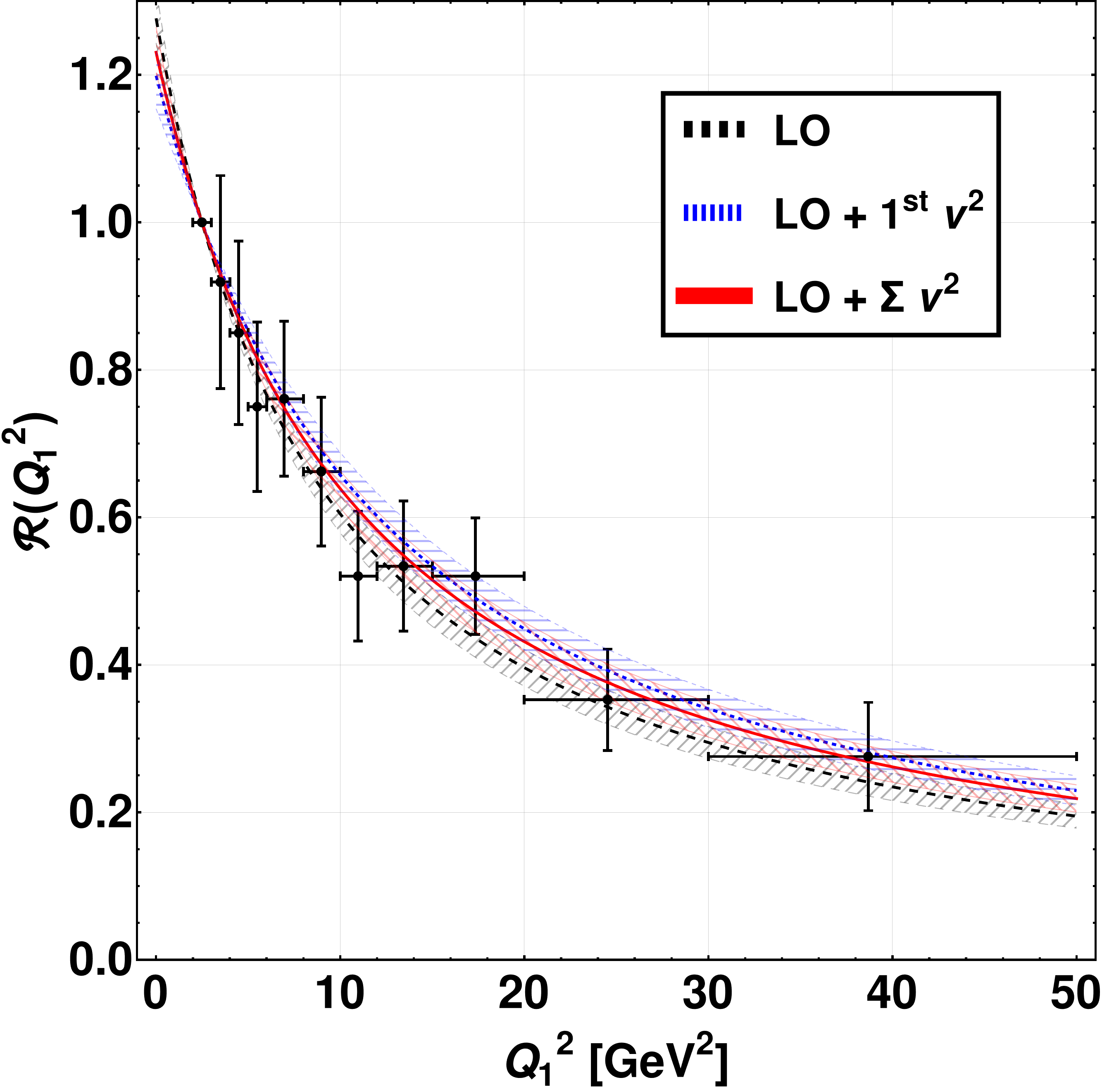} \label{fig:singleplotsrelativisticRe}}
    \subfloat[]{\includegraphics[width=5.5cm]{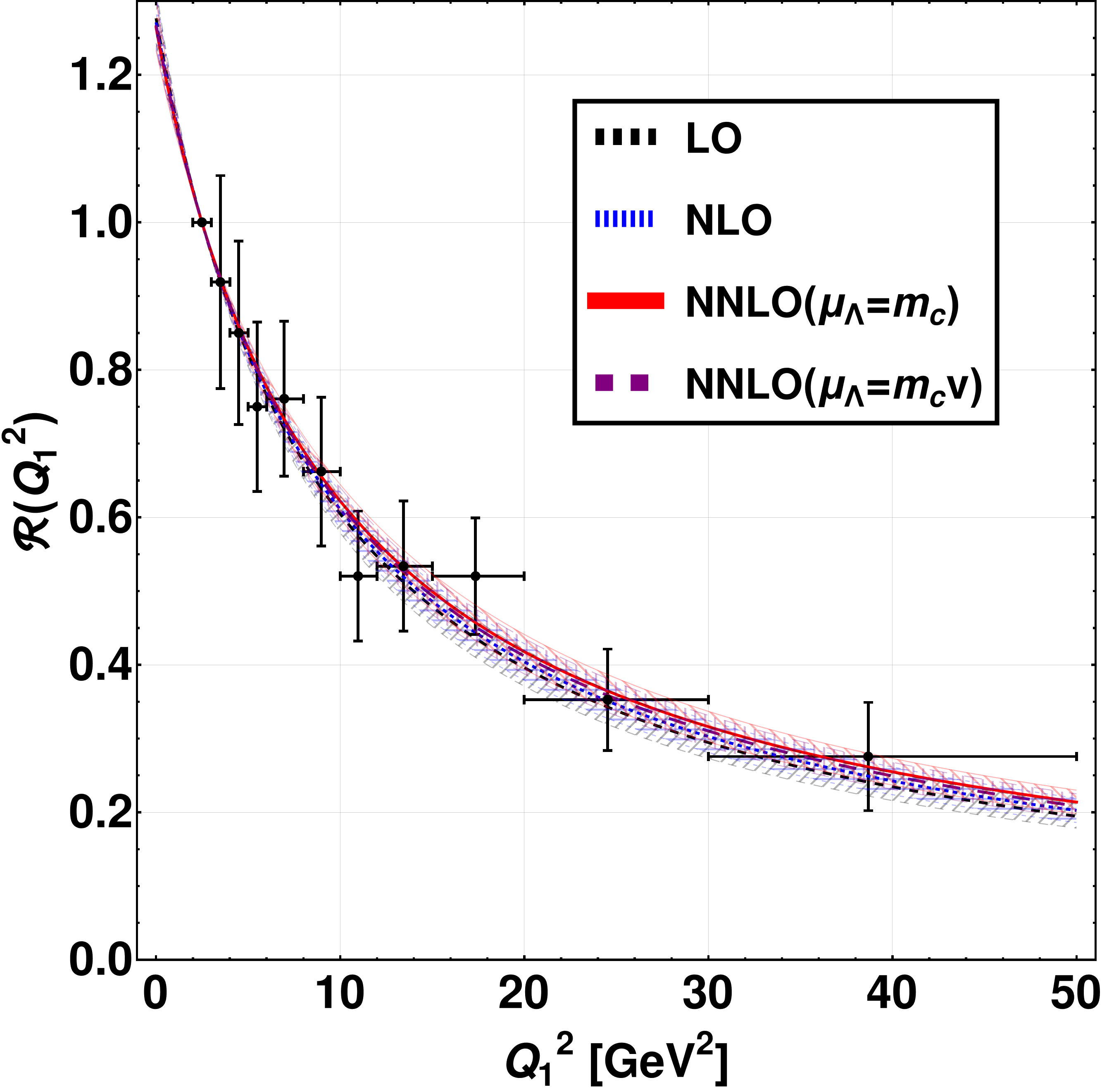} \label{fig:singleplotsQCDRe}}
    \subfloat[]{\includegraphics[width=5.5cm]{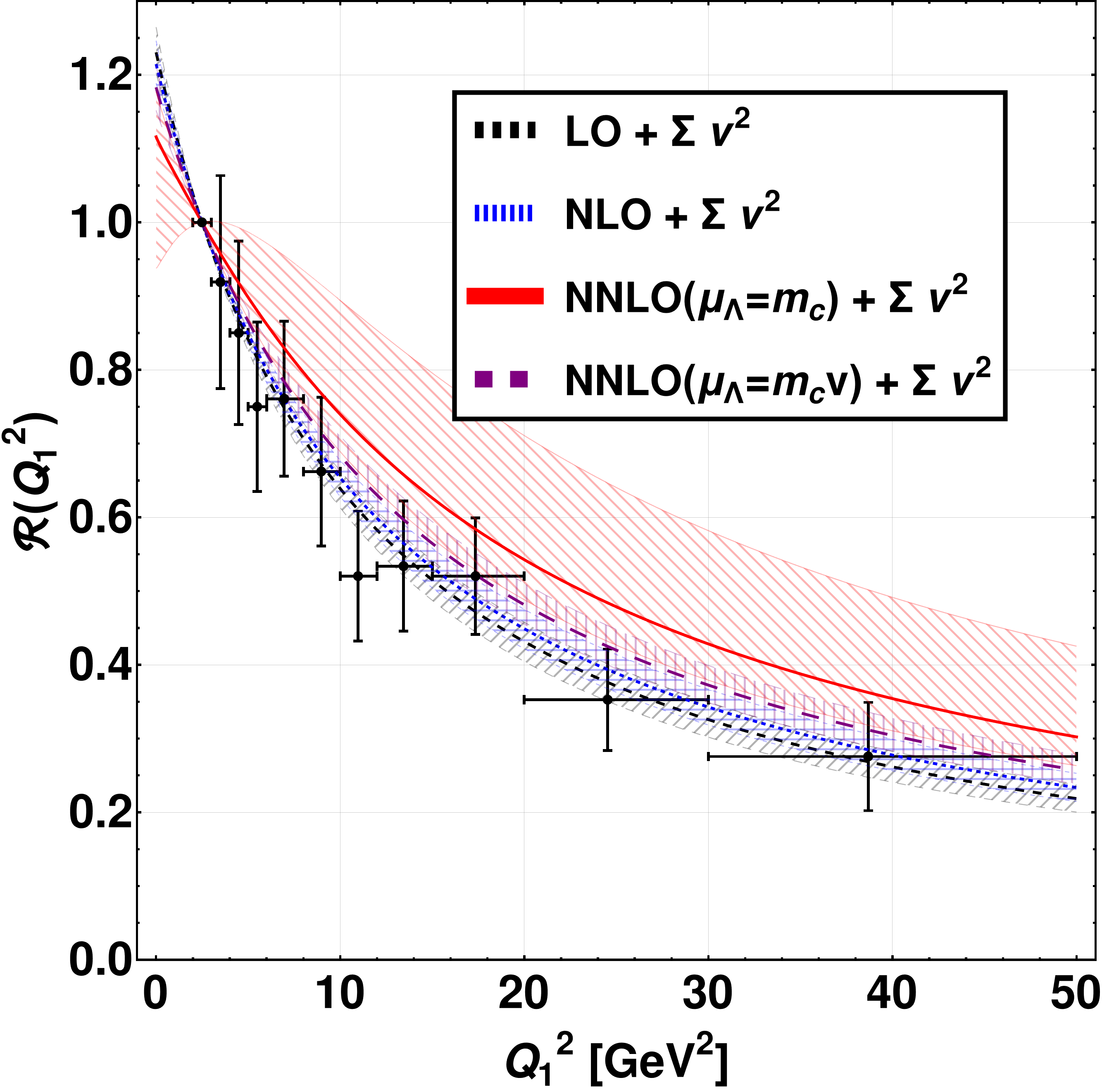} \label{fig:singleplotsQCDrelativisticRe}}
  \end{center}
  \caption{Transition form-factor ratio $\mathcal{R}{(Q_1^2)}$ with impact of
  (a) relativistic corrections, (b) QCD corrections, 
  (c) combined relativistic and QCD corrections.}
  \label{fig:singleplotsRe}
\end{figure}

We have exhibited in Fig.~\ref{fig:singleplotsRe}, in the same fashion as before, the pure relativistic, the pure QCD and the combined QCD and relativistic corrections. As can be seen in Fig.~\ref{fig:singleplotsrelativisticRe}, we note that, unlike in the previous case, the pure relativistic corrections are able to describe the data well. Similarly as before, the pure NNLO contributions in Fig.~\ref{fig:singleplotsQCDRe} are in agreement with the data as well. In the case of the combined QCD and relativistic corrections in Fig.~\ref{fig:singleplotsQCDrelativisticRe}, we note that this time the combined NNLO correction with all-order $v^2$ resummation at the NRQCD scale $\mu_{\Lambda}=m_c \sqrt{\langle v^2 \rangle_{\eta_c}}$ is largely compatible with the \textsc{Ba}\textsc{Bar} data.

\subsection{Double space-like transition form-factor}

In this subsection, we discuss corrections to observables in the process $ \gamma^* \gamma^* \rightarrow \eta_c$ with two off-shell photons. These observables could be measured, in the future, at lepton colliders by tagging both leptons. In the following, we present, for the first time, the NNLO corrections in the strong coupling constant for the double off-shell case. We will also discuss the impact of the relativistic corrections.

For this, we define the observable
\begin{equation}
\begin{split}
    R{(Q_1^2, Q_2^2)}=&\frac{\left\vert \mathcal{F}_{\eta_c}(t_1,t_2) \right\vert}{\left\vert \mathcal{F}_{\eta_c}(0,0) \right\vert}  
    \\
    =& \frac{4}{4 - t_1 - t_2} \frac{\left\vert 1 + g(t_1, t_2, \langle v^2 \rangle_{\eta_c}) + \Delta(t_1, t_2) \right\vert}{\left\vert 1 + g(0,0, \langle v^2 \rangle_{\eta_c}) + \Delta(0, 0) \right\vert} \, ,
\end{split}
\label{eq:doubleobservR}
\end{equation}
where the virtualities of the two photons are defined as
\begin{equation}
    Q_1^2=-m_c^2\, t_1 \, , \qquad\qquad Q_2^2=-m_c^2\, t_2 \, .
\end{equation}
In the limit, where one of the photons becomes on-shell, eq.~(\ref{eq:doubleobservR}) becomes equivalent to the observable in eq.~(\ref{eq:Rsingleobservable}) from the previous subsection. We proceed to compute the central values and the uncertainties in a similar fashion as before.

\begin{figure}
    \begin{center}
    \subfloat[]{\includegraphics[width=5.5cm]{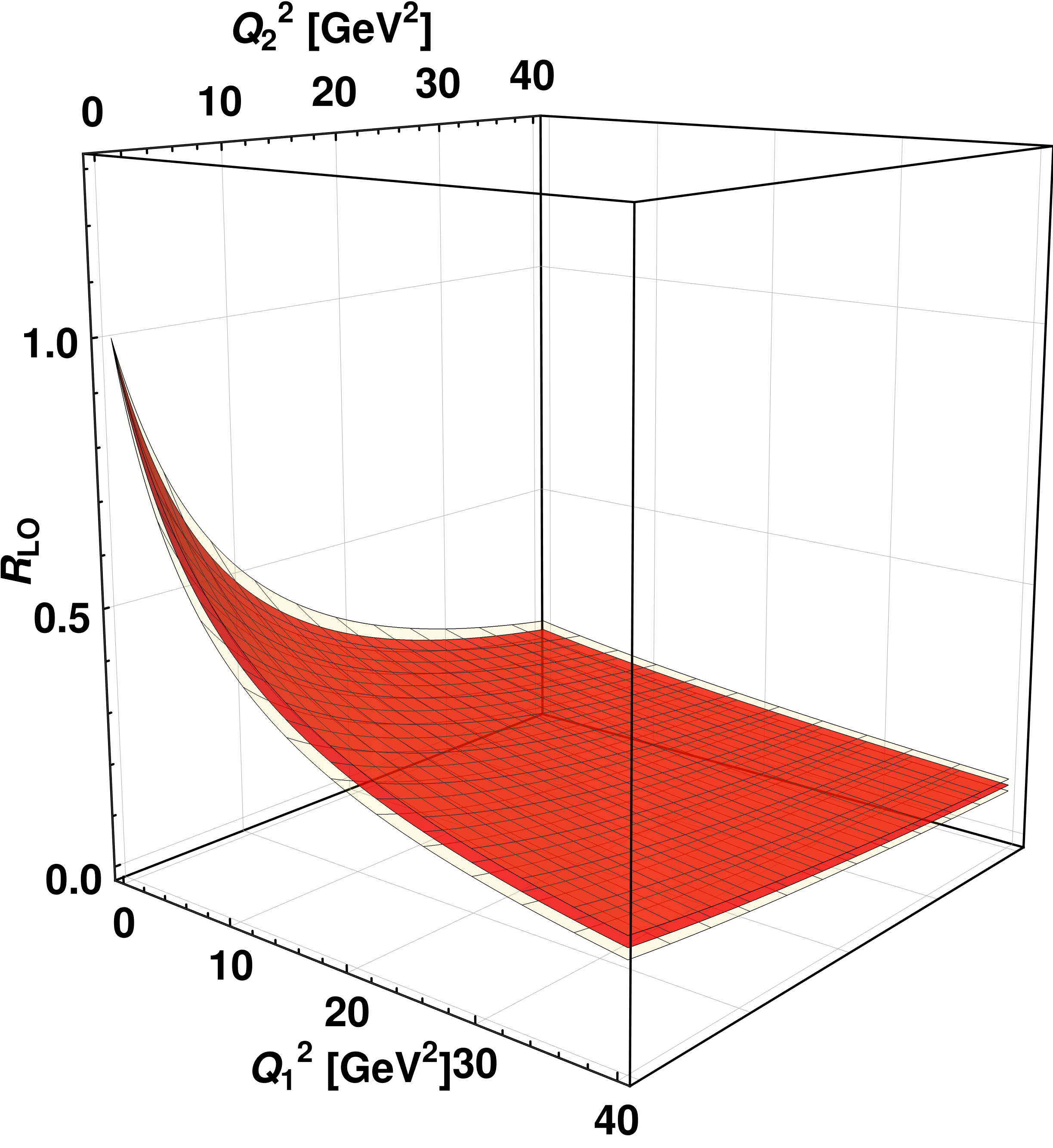} \label{fig:RDoubleQ1Q2LO}}
    \subfloat[]{\includegraphics[width=5.5cm]{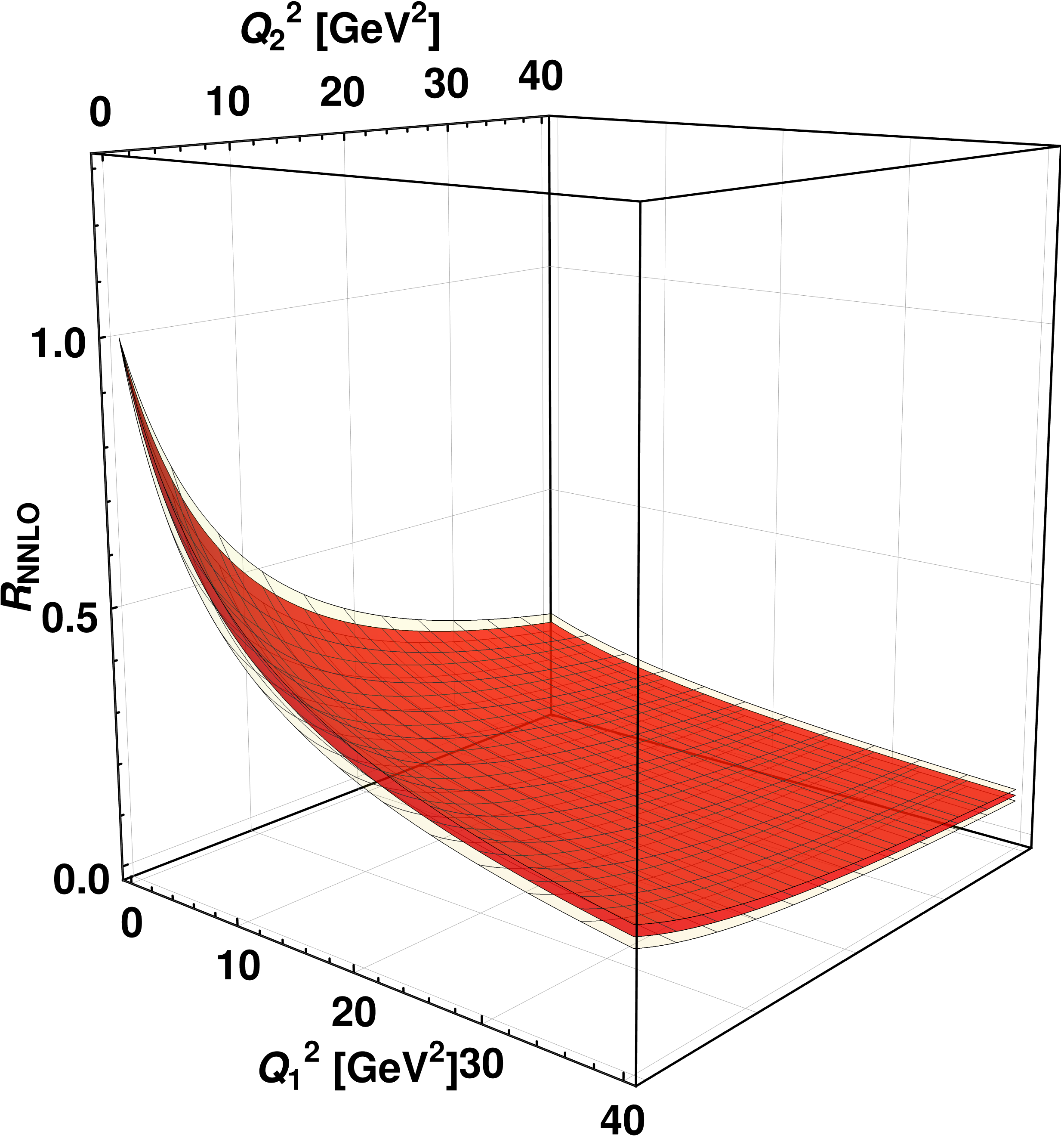} \label{fig:RDoubleQ1Q2NNLO}}
    \subfloat[]{\includegraphics[width=5.5cm]{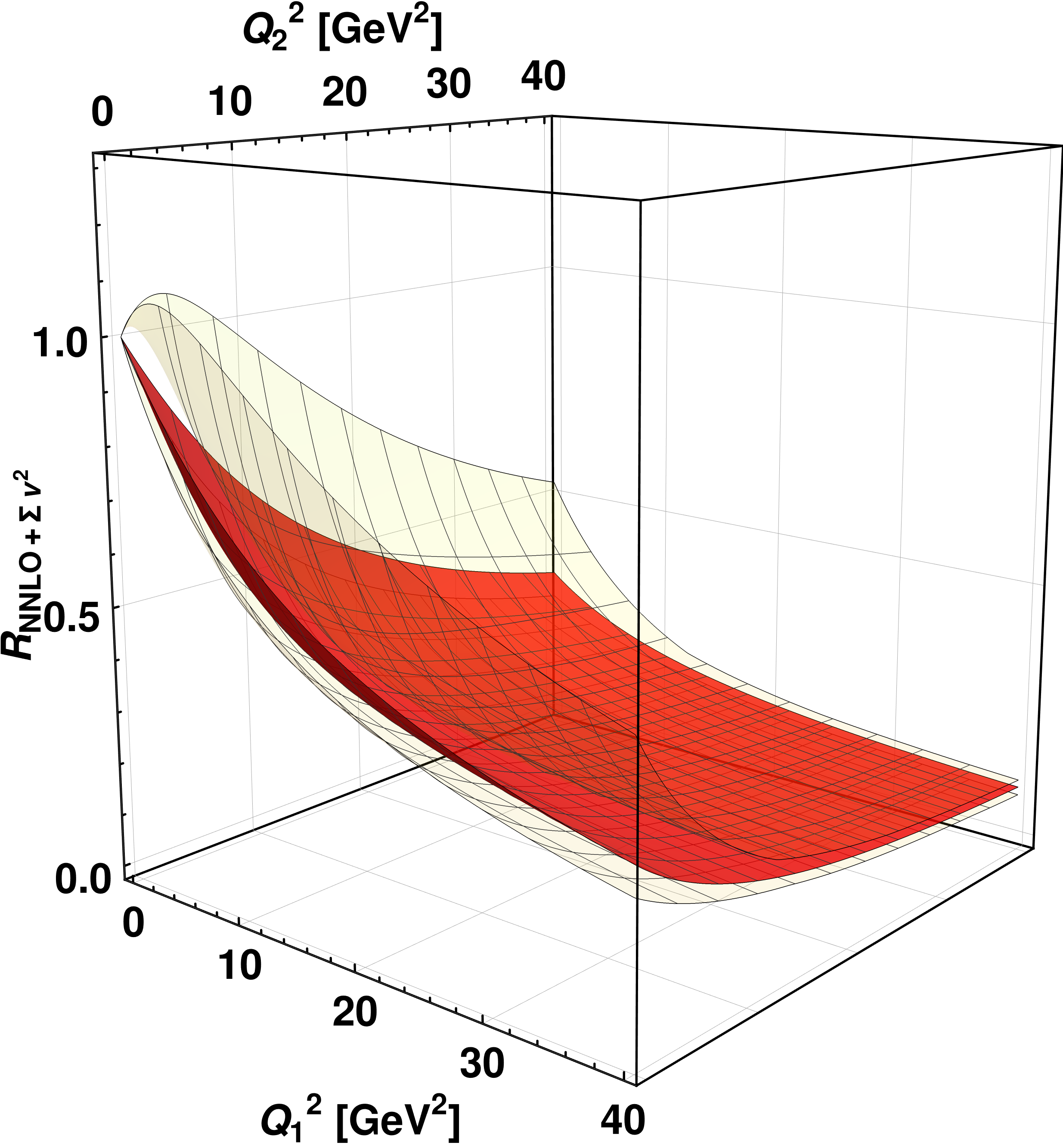} \label{fig:RDoubleQ1Q2NNLOallvsq}}
    \end{center}
    \caption{The ratio of the transition form-factors, $R{(Q_1^2, Q_2^2)}$ (see eq.~(\ref{eq:doubleobservR})), for (a) LO, (b) NNLO ($\mu_{\Lambda}=m_c$) and (c) NNLO ($\mu_{\Lambda}=m_c$) $+ \sum v^2$ as a function of $Q_1^2$ and $Q_2^2$. The red plane represents the central value and the two transparent yellow planes correspond to the upper and lower bounds.}
    \label{fig:RDoubleQ1Q2}
\end{figure}

In Fig.~\ref{fig:RDoubleQ1Q2}, we display the normalised transition form-factor, $R{(Q_1^2, Q_2^2)}$, with the central values and their uncertainties as a function of $Q_1^2$ and $Q_2^2$ at LO, at NNLO, and at NNLO~$+ \sum v^2$. For the NRQCD scale, we set $\mu_{\Lambda}=m_c$.

We observe that, while the pure NNLO corrections in Fig.~\ref{fig:RDoubleQ1Q2NNLO} follow well the shape of the LO plane in Fig.~\ref{fig:RDoubleQ1Q2LO}, the shape is more pronounced for the case of the combined NNLO and relativistic corrections in Fig.~\ref{fig:RDoubleQ1Q2NNLOallvsq}. This can be traced back to the fact that the correction at the on-shell point leads to a small value, hence induces large variation in the ratio in eq.~(\ref{eq:doubleobservR}).

In addition, we also consider the following observable on the $\tau$ and $\omega$ plane (see eq.~(\ref{eq:tauomegadef})),
\begin{equation}
\begin{split}
    T{(\tau, \omega)}=&\frac{\left\vert \mathcal{F}_{\eta_c}(t_1,t_2) \right\vert}{\left\vert \mathcal{F}_{\eta_c}(\bar{t},\bar{t}) \right\vert}  
    \\
    =& \frac{\left\vert 1 + g(t_1, t_2, \langle v^2 \rangle_{\eta_c}) + \Delta(t_1, t_2) \right\vert}{\left\vert 1 + g(\bar{t},\bar{t}, \langle v^2 \rangle_{\eta_c}) + \Delta(\bar{t}, \bar{t}) \right\vert} \, ,
\end{split}
\label{eq:doubleobservtauomega}
\end{equation}
where
\begin{equation}
    t_1 = -2 (1+\omega) \tau \, , \qquad t_2 = -2 (1-\omega) \tau \, , \qquad \bar{t} = -2 \tau \, .
\end{equation}
For this observable, we have trivially that, at LO, $T{(\tau, \omega)}=1$. We will investigate the impact of the different corrections as a function of $\omega$ and for fixed $\tau$.

\begin{figure}
    \begin{center}
    \subfloat[]{\includegraphics[width=5.62cm]{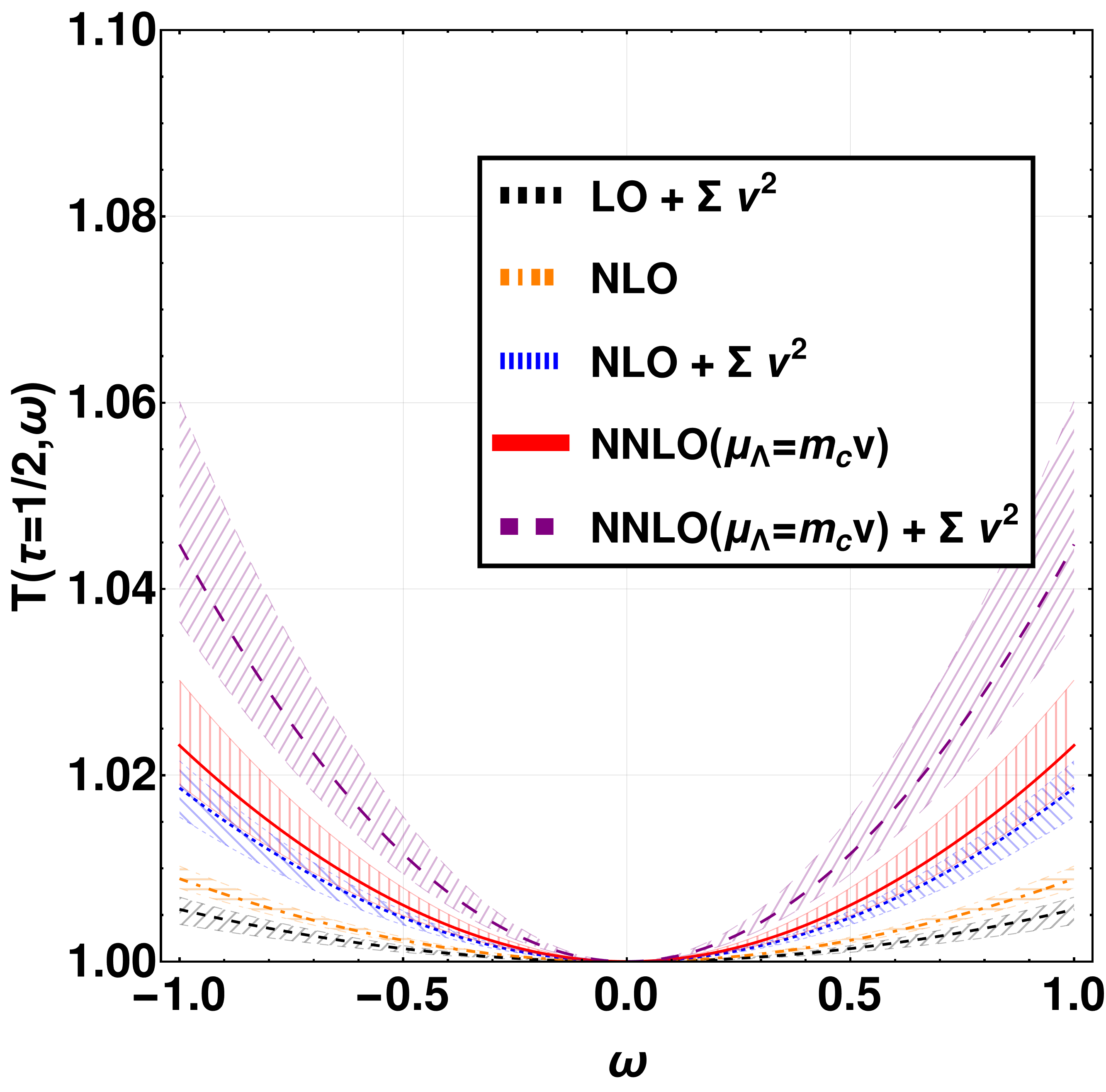} \label{fig:Ttauomegahalf}}
    \subfloat[]{\includegraphics[width=5.62cm]{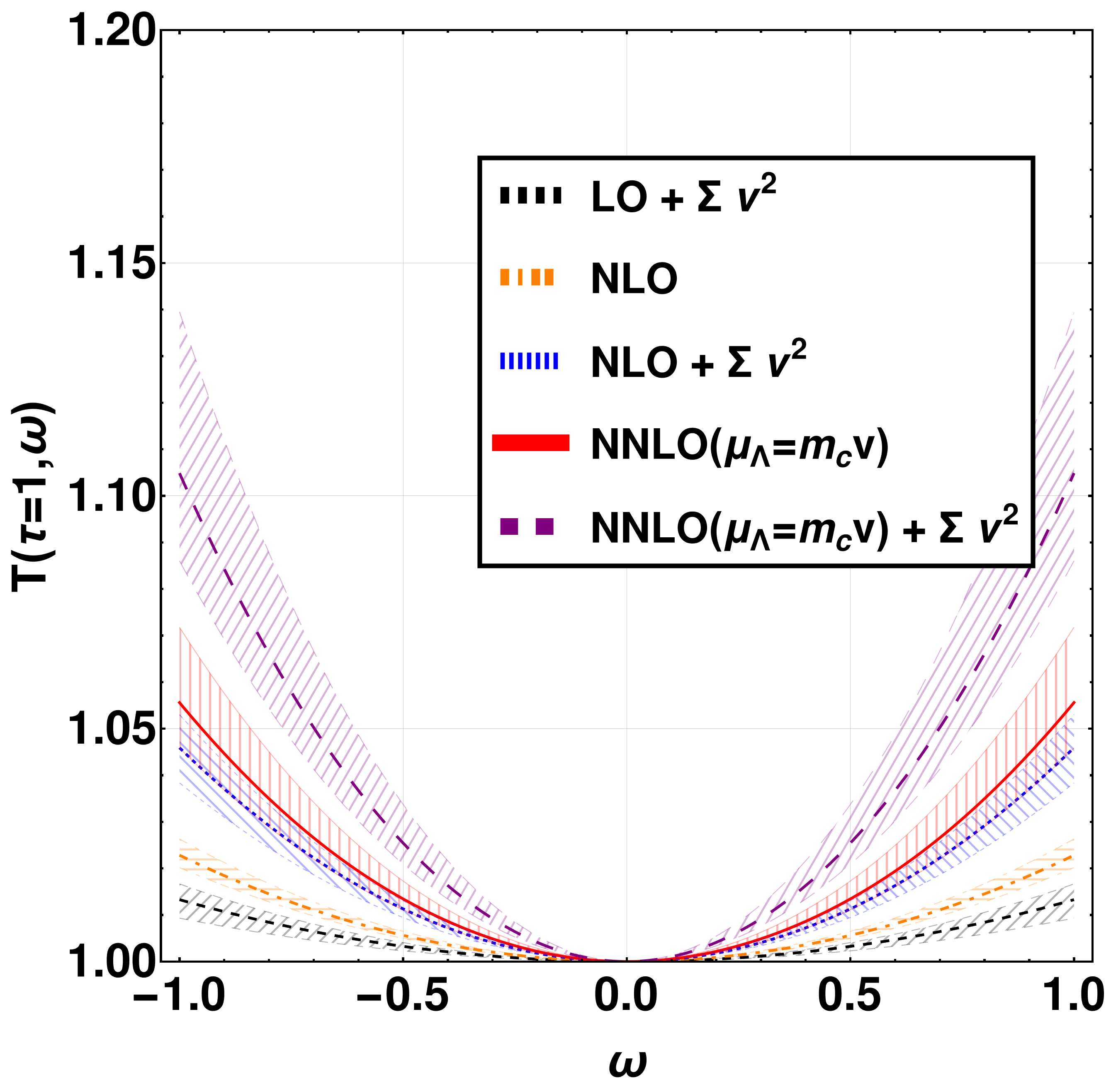} \label{fig:Ttauomegaone}}
    \subfloat[]{\includegraphics[width=5.5cm]{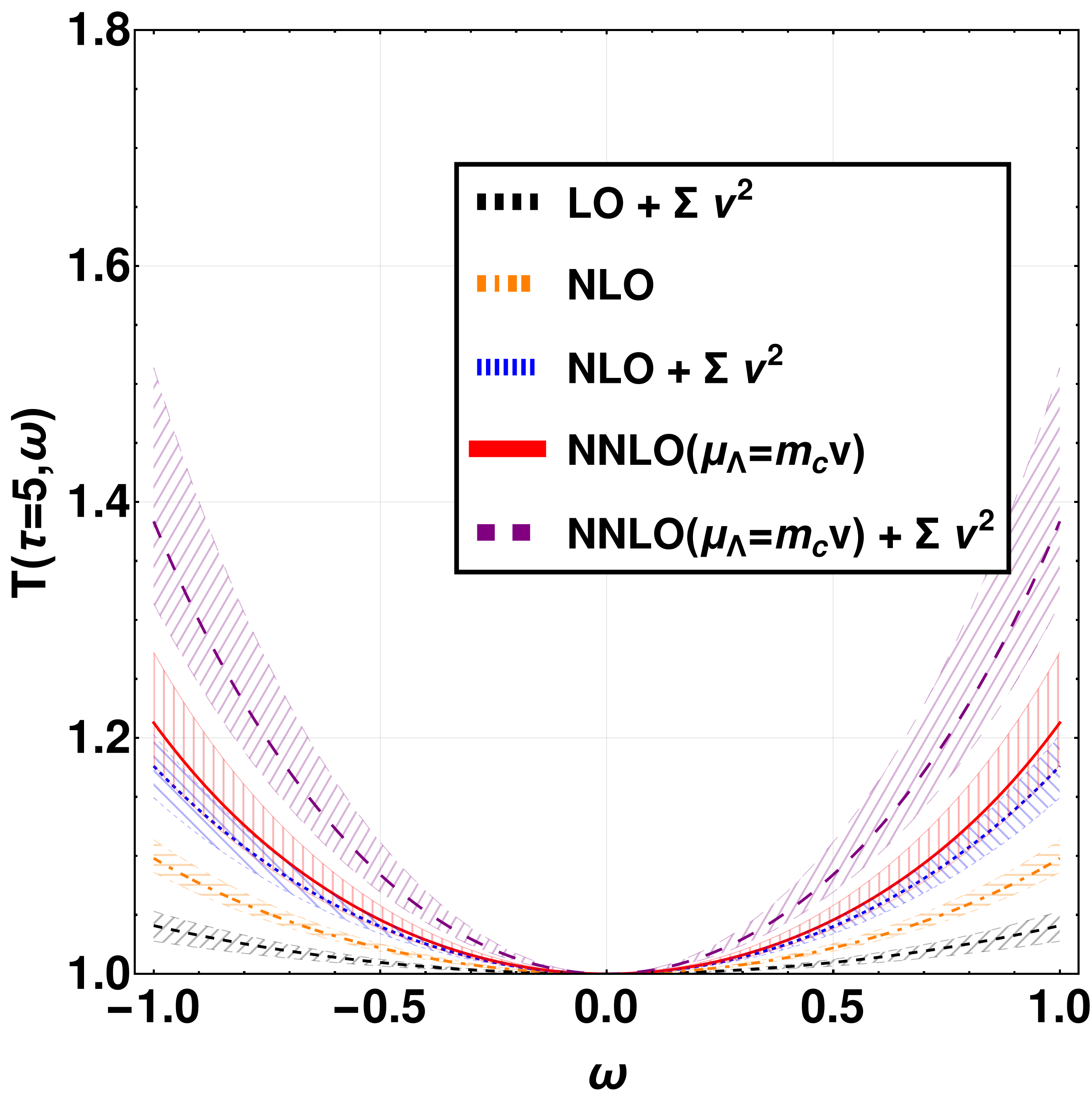} \label{fig:Ttauomegafive}}
    \end{center}
    \caption{The impact of the different corrections on the observable $T{(\tau, \omega)}$ (see eq.~(\ref{eq:doubleobservtauomega})) as a function of $\omega$ at (a) $\tau=1/2$, (b) $\tau=1$ and (c) $\tau=5$.}
    \label{fig:Ttauomega}
\end{figure}

In Fig.~\ref{fig:Ttauomega}, we show the effect of the different corrections at $\tau=1/2$, $\tau=1$ and $\tau=5$. For the NNLO corrections, we show the curves this time with the NRQCD scale $\mu_{\Lambda}=m_c \sqrt{\langle v^2 \rangle_{\eta_c}}$.

While, by definition, all corrections are equal to unity at $\omega=0$, the corrections increase as one moves towards the two endpoints, $\omega=\pm 1$. We also note that the relative size of the corrections increases for larger $\tau$. However, the overall shape along the $\omega$ axis remains that of a symmetric parabola centered at $\omega=0$.

While the pure relativistic correction is rather small, the size of the pure NLO correction can reach up to $10\%$ at $\omega=\pm 1$ for $\tau=5$. Combining the NLO and relativistic corrections nearly doubles the size of the pure NLO correction. We note that the pure NNLO correction is, for the whole $\omega$ range, slightly larger than the combined NLO $+ \sum v^2$ part. At $\tau=1/2$, it can reach up to $3\%$ at the boundaries, while at $\tau=1$, it reaches up to $7\%$, and for large $\tau=5$, the size is just over $20\%$. Combining NNLO and relativistic corrections has a similar effect as in the NLO case, where the size almost doubles compared to the pure QCD correction case.

We anticipate that, in the future, experiments such as Belle II and the upcoming Future Circular Collider (FCC) should, in principle, be able to perform measurements of the $\eta_c$ transition form-factor with two off-shell photons. This requires cross-section measurements with two tagged electrons. Our theory predictions can then be used to compare against the experimental result.

\subsection{Decay width for $\eta_c \rightarrow \gamma \gamma$}

In this subsection, we briefly discuss the impact of the different corrections on the $\eta_c$ decay width to two photons. In the previous two subsections, where we considered ratios of transition form-factors, the LDME $\bra{\eta_c} | \psi^{\dag} \chi | \ket{0}$ cancelled out. However, for the observable here, we need to include it and, using eq.~(\ref{eq:LDMER0}) to express it in terms of $R_{\eta_c}{(0)}$, we have that 
\begin{equation}
\begin{split}
{\cal F}_{\eta_c}(t_1, t_2) = \frac{e_c^2 \sqrt{N_c} \, R_{\eta_c}{(0)}}{\sqrt{2 \pi m_c^5}} \,  \frac{\big(1+\langle v^2 \rangle_{\eta_c}\big)^{\frac{1}{4}}}{1+\tau}  \Big ( 1 + g(t_1,t_2, \langle v^2 \rangle_{\eta_c}) + \Delta(t_1, t_2)\Big) \, .
\end{split}
\end{equation}
For the decay width to $\gamma \gamma$, we need the form-factor at the double on-shell point, $t_1=t_2=0$, and obtain
\begin{equation}
\begin{split}
    {\cal F}_{\eta_c}(0,0) =& \frac{e_c^2 \sqrt{N_c}\, R_{\eta_c}{(0)}}{\sqrt{2 \pi m_c^5}} \, \Bigg[ \frac{1}{2 \sqrt{\langle v^2 \rangle_{\eta_c}} \big(1+\langle v^2 \rangle_{\eta_c}\big)^{\frac{5}{4}}} \log \Bigg[ \frac{ \sqrt{1 + \langle v^2 \rangle_{\eta_c}} + \sqrt{\langle v^2 \rangle_{\eta_c}} }{\sqrt{1 + \langle v^2 \rangle_{\eta_c}} - \sqrt{\langle v^2 \rangle_{\eta_c}}} \Bigg]
    \\
    &\hspace{2.4cm} + \big(1+\langle v^2 \rangle_{\eta_c}\big)^{\frac{1}{4}}\, \Delta(0,0)\Bigg]\, ,
\end{split}
\end{equation}
where we used eq.~(\ref{eq:allorderrelativistict10t20}) to write out $g(0,0,\langle v^2 \rangle_{\eta_c})$ explicitly. Using this, the decay width to di-photon can now be written as
\begin{align}
    \Gamma[\eta_c \rightarrow \gamma \gamma] =& \frac{\pi \alpha^2 M_{\eta_c}^3}{4} \left\vert {\cal F}_{\eta_c}(0,0)\right\vert^2
    \nonumber \\
    =& \frac{\alpha^2 e_c^4 N_c \, |R_{\eta_c}{(0)}|^2}{m_c^2} \, \Bigg[ \frac{1}{4 \langle v^2 \rangle_{\eta_c} \big(1+\langle v^2 \rangle_{\eta_c}\big)} \log^2 \Bigg[ \frac{ \sqrt{1 + \langle v^2 \rangle_{\eta_c}} + \sqrt{\langle v^2 \rangle_{\eta_c}} }{\sqrt{1 + \langle v^2 \rangle_{\eta_c}} - \sqrt{\langle v^2 \rangle_{\eta_c}}} \Bigg]
    \nonumber \\
    &\hspace{2.4cm} + \frac{\sqrt{1+\langle v^2 \rangle_{\eta_c}}}{\sqrt{\langle v^2 \rangle_{\eta_c}}} \log \Bigg[ \frac{ \sqrt{1 + \langle v^2 \rangle_{\eta_c}} + \sqrt{\langle v^2 \rangle_{\eta_c}} }{\sqrt{1 + \langle v^2 \rangle_{\eta_c}} - \sqrt{\langle v^2 \rangle_{\eta_c}}} \Bigg] \Ree{[\Delta(0,0)]}
    \nonumber \\
    &\hspace{2.4cm} + \big(1+\langle v^2 \rangle_{\eta_c}\big)^2 \, |\Delta(0,0)|^2 \Bigg]\, ,
\label{eq:Gamma_etac}
\end{align}
where we have taken the full square of the form-factor and $\Ree{[\Delta(0,0)]}$ stands for the real part of $\Delta(0,0)$. Expanding the decay width to second order in $\langle v^2 \rangle_{\eta_c}$, we obtain
\begin{equation}
\begin{split}
    \Gamma[\eta_c \rightarrow \gamma \gamma] =& \frac{\alpha^2 e_c^4 N_c \, |R_{\eta_c}{(0)}|^2}{m_c^2} \, \left( 1 -\frac{4}{3}  \langle v^2 \rangle_{\eta_c} +  \frac{68}{45} \langle v^2 \rangle^2_{\eta_c} + \dots \right)\, ,
\end{split}
\end{equation}
where the coefficients of the first two relativistic corrections agree with the results derived previously in ref.~\cite{Bodwin:1994jh,Keung:1982jb,Bodwin:2002cfe}.

\begin{table}[h]
    \centering
    \begin{tabular}{| l || c | c |}
    \hline
            & $\Gamma[\eta_c \rightarrow \gamma \gamma]$~[keV] & $|R_{\eta_c}{(0)}|^2_{\rm fit}$~[$\text{GeV}^3$]  \\
      \hline \hline
      LO                     &  $12.9^{+3.2}_{-3.1}$ & $0.36^{+0.06}_{-0.05}$ \\
      LO + $1^{\rm st}$ $v^2$   & $9.5^{+2.4}_{-2.3}$ & $0.50^{+0.08}_{-0.07}$  \\
      LO + $\sum v^2$   & $10.1^{+2.5}_{-2.4}$ & $0.46^{+0.08}_{-0.07}$ \\
      NLO                    & $9.9^{+2.5}_{-2.4}$ & $0.47^{+0.08}_{-0.07}$ \\
      NLO + $\sum v^2$  & $7.0^{+1.7}_{-1.7}$ & $0.67^{+0.11}_{-0.10}$ \\
      NNLO ($\mu_0=\mu_{\Lambda}=m_c$)    & $3.1^{+1.2}_{-1.5}$ & $1.53^{+0.63}_{-0.74}$ \\
      NNLO ($\mu_0=\mu_{\Lambda}=m_c v$)    & $5.8^{+1.5}_{-1.4}$ & $0.81^{+0.14}_{-0.17}$ \\
      NNLO ($\mu_0=\mu_{\Lambda}=m_c$) + $\sum v^2$  & $0.9^{+0.9}_{-0.8}$ & $4.99^{+4.83}_{-4.20}$ \\
      NNLO ($\mu_0=\mu_{\Lambda}=m_c v$) + $\sum v^2$  & $3.1^{+0.8}_{-1.0}$ & $1.51^{+0.40}_{-0.48}$ \\
      \hline \hline
      PDG                    & $5.1 \pm 0.4$ & \\
      \hline
    \end{tabular}
    \caption{Radiative decay width of $\eta_c$ to di-photon at different orders in $\alpha_s$ and $\langle v^2 \rangle_{\eta_c}$ in the middle column using the value of $|R_{\eta_c}{(0)}|^2$ in eq.~(\ref{eq:R0standdef}), and fit of $|R_{\eta_c}{(0)}|^2_{\rm fit}$ in the right column to the experimental PDG value \cite{ParticleDataGroup:2024cfk}.
    }
    \label{tab:decaywidth}
\end{table}

We have collected, in Table~\ref{tab:decaywidth}, the values of the decay width $\Gamma[\eta_c \rightarrow \gamma \gamma]$ at different orders in perturbative QCD and the relativistic expansion. The uncertainties are computed according to the procedure explained at the beginning of Section~\ref{Sec:Applications}. We observe that, while the LO decay width exceeds the PDG value \cite{ParticleDataGroup:2024cfk} by more than a factor of two, both relativistic and QCD corrections are negative and decrease this value.

We also note that the relativistic corrections are of similar size as the pure NLO correction. As for the NNLO corrections, similarly as in the previous subsections, we investigate the impact at different values for $\mu_{\Lambda}$, where we set $\mu_0=\mu_{\Lambda}$ for the wavefunction at the origin. We find that NNLO corrections, in particular those at $\mu_{\Lambda}=m_c \sqrt{\langle v^2 \rangle_{\eta_c}}$ are closest to the PDG value \cite{ParticleDataGroup:2024cfk}. Adding relativistic corrections pushes the value further down.

As outlined at the beginning of Section~\ref{Sec:Applications}, the value for the wavefunction at the origin, $|R_{\eta_c}{(0)}|^2$, in ref.~\cite{Bodwin:2007fz} was extracted by a fit to the old PDG value of the di-photon decay width at $\Gamma[\eta_c \rightarrow \gamma \gamma]_{\text{old}} = \left( 7.2 \pm 2.1 \right) \text{ keV}$ \cite{ParticleDataGroup:2006fqo}. In principle, it is possible to extract a value for $|R_{\eta_c}{(0)}|^2$ at each order in $\alpha_s$ and in $\langle v^2 \rangle_{\eta_c}$. We have done so by performing a direct fit to the current PDG value \cite{ParticleDataGroup:2024cfk}, $\Gamma[\eta_c \rightarrow \gamma \gamma]_{\text{PDG}} = \left( 5.1 \pm 0.4 \right) \text{ keV}$, using eq.~(\ref{eq:Gamma_etac}). The result for $|R_{\eta_c}{(0)}|^2_{\rm fit}$ at the different orders is given in the right column of Table~\ref{tab:decaywidth}. It remains to be seen whether, in addition, a fit to $\langle v^2 \rangle_{\eta_c}$ can also be performed at each order in $\alpha_s$.

\section{Conclusions}
\label{Sec:Conclusions}

In this paper, we have studied both relativistic and perturbative QCD corrections to the transition form-factor ${\cal F}_{\eta_Q}{(t_1,t_2)}$ for the process $\gamma^* \gamma^* \leftrightarrow \eta_Q$ as a function of the two normalised photon virtualities $t_1$ and $t_2$. On the relativistic side, we have resummed a class of relativistic corrections up to all orders in $v^2$, while on the perturbative QCD side, we have computed QCD corrections up to Next-to-Next-to-Leading Order in the strong coupling $\alpha_s$. While the two-loop amplitudes were computed before in the case of double on-shell and the case of a single off-shell photon, these were previously unknown for the case of double off-shell photons considered in this work.

As such, the two-loop correction factor depends on the two normalised photon virtualities, $t_i$, and can be expressed in terms of a set of previously unknown master integrals that we have computed numerically in this paper. We have cross-checked our results in the corresponding kinematic limits and were able to reproduce the known results in the literature. The two-loop amplitudes obtained in this work can be used to compute, in addition to the NNLO QCD corrections for pseudo-scalar quarkonia, also the NNLO QED corrections for para-leptonium states. 

Having at disposal the relativistic and QCD corrections, we have then explored three different phenomenological applications for $\eta_c$. We first discussed the case with one off-shell photon and compared our predictions for the observable $R{(Q_1^2)}$ with the existing $\eta_c$ measurements at \textsc{Ba}\textsc{Bar} \cite{BaBar:2010siw}. We establish that, while the NNLO QCD corrections are able to describe the data quite well, combining QCD with relativistic corrections slightly overshoot the data. In addition, we have analysed the alternative observable $\mathcal{R}{(Q_1^2)}$, that eliminates the detection efficiency dependence of the non-tagged cross-section, and find that the combined QCD and relativistic corrections have better agreement with the data than before.

For the second phenomenological application, we then investigated the transition form-factor for the double off-shell case, $R{(Q_1^2, Q_2^2)}$, as a function of the two virtualities $Q_1^2$ and $Q_2^2$. We observe that, while the NNLO corrections follow well the shape of the LO contribution, the combined NNLO and relativistic corrections can induce larger variations. Instead of normalising to the double on-shell point, we then also computed the observable $T{(\tau, \omega)}$ as a function of $\omega$ at fixed $\tau$. We provided theory predictions using both relativistic and QCD corrections. We hope that, in the future, experimentalists will be in a position to measure the $\eta_c$ transition form-factor with dependencies on both photon virtualities.

Finally, we revisited the decay width of $\eta_c$ to di-photon and computed the observable at different orders in $\alpha_s$ and $\langle v^2 \rangle_{\eta_c}$. Comparing it with the current PDG value \cite{ParticleDataGroup:2024cfk}, we determine that, while the LO prediction exceeds the PDG value by more than a factor of two, the pure NNLO corrections are consistent with it. In contrast to this, the combined QCD and relativistic corrections undershoot the data slightly. Having performed these predictions at different orders in $\alpha_s$ with the same $|R_{\eta_c}{(0)}|^2$, we have then extracted the corresponding values of the wavefunction by fitting directly to the PDG value.

We conclude by stating that, while the pure NNLO corrections are able to describe the existing data well, combining QCD and relativistic corrections can induce larger uncertainties. In view of this, it will be intriguing to see whether a simultaneous fit to both $|R_{\eta_c}{(0)}|^2$ and $\langle v^2 \rangle_{\eta_c}$ can be performed at each order in $\alpha_s$. In addition, it would be interesting to survey the impact of the mixed QCD and relativistic corrections at $\mathcal{O}{(\alpha_s v^2)}$. We leave this for future work.

\acknowledgments{We thank G.T.~Bodwin, V.~Druzhinin, and R.~Mizuk for useful discussions. We are indebted to IFJ PAN and IJCLab for support in the framework of the IFJ PAN-IJCLab collaboration. This work was partially supported by the Polish National Science Center Grant No. UMO-2023/49/B/ST2/03665 and by the French National Research Agency via the ANR PIA Grant No. ANR-20-IDEES-0002.}

\appendix

\section{Resummation of $v^2$ corrections}
\label{sec:appendixresummation}

In this appendix, we briefly outline the procedure to resum a certain class of relativistic corrections. We follow the procedure in Bodwin et al. in ref.~\cite{Bodwin:2007fz} who have suggested a way to resum relativistic corrections that stem from operators as in
\begin{equation}
\Bigbra{\eta_Q} \Big| \psi^{\dag} 
\Big(-i \frac{\tensor{\mathbf D}}{2} \Big)^{2n}  \chi \Big| \Bigket{0} = \langle |\vec k|^{2n} \rangle_{\eta_Q} \,  
\Bigbra{\eta_Q} \Big| \psi^{\dag} \chi \Big| \Bigket{0} \, .
\end{equation}
To this end, we start from the reduced form-factor in eq.~(\ref{eq:FF_sd_all_order}), that includes the full dependence on $|\vec k|/m_Q$, 
\begin{equation}
 {\cal F}_{^1S_0}(t_1,t_2,\vert \vec k \vert^2) =  -e_Q^2 \sqrt{2N_c} \, \frac{m_Q}{E} \tilde{A_0}.
\end{equation}
We have made its dependence on $\vert \vec k \vert^2$ explicit, and expand it in terms of short-distance coefficients $c_n(t_1,t_2)$:
\begin{equation}
\begin{split}
    {\cal F}_{^1S_0}(t_1,t_2, \vert \vec k \vert^2) =& \sum_{n} c_n(t_1,t_2) \Big \langle (Q \bar Q)[^1S^{[1]}_0] \Big| \psi^{\dag} \Big(-i \frac{\tensor{{\mathbf D}}}{2} \Big)^{2n} \chi \Big| 0 \Big \rangle
    \\
    =& \sqrt{2 N_c}\, 2E \sum_n c_n(t_1,t_2) \,  \vert\vec{k}\vert^{2n} \, ,
    \label{eq:FF_pQCD_all_order}
\end{split}
\end{equation}
where we made use of
\begin{equation}
      \Bigbra{(Q\bar Q)[^1S^{[1]}_0]} \Big| \psi^{\dag} \Big(-i \frac{\tensor{\mathbf D}}{2} \Big)^{2n} \chi \Big| \Bigket{0 } = \sqrt{2N_c}\, 2E \, \vert\vec{k}\vert^{2n} \, .
\end{equation}
We can now read off the short-distance coefficients as
\begin{equation}
c_n(t_1,t_2) = \frac{1}{2 \sqrt{2 N_c}} \, \frac{1}{n!} \, \Big( \frac{\partial^n}{\partial \vert \vec k\vert^{2n}}\, 
\frac{ {\cal F}_{^1S_0}(t_1,t_2,\vert \vec k \vert^2)}{E(\vert \vec k \vert^2)} \Big)\Big|_{\vert \vec k \vert^2 = 0} \, ,
\end{equation}
where we made the dependence of $E$ on $\vert \vec k \vert^2$ explicit. The hadronic form-factor on the other hand is
\begin{equation}
\begin{split}
    {\cal F}_{\eta_Q}(t_1,t_2) =& \sum_{n} c_n(t_1,t_2) \Big \langle \eta_Q \Big| \psi^{\dag} \Big(-i \frac{\tensor{{\mathbf D}}}{2} \Big)^{2n} \chi \Big| 0 \Big \rangle
    \\
    =& \Bigbra{\eta_Q} \Big| \psi^{\dag} \chi \Big| \Bigket{0} \sum_n c_n(t_1,t_2) \, \langle \vert\vec k\vert^{2n} \rangle_{\eta_Q} \, .
\end{split}
\end{equation}
If now, following ref.~\cite{Bodwin:2007fz}, we replace
\begin{equation}
   \langle \vert\vec k\vert^{2n} \rangle_{\eta_Q} = \langle \vert\vec k\vert^2  \rangle^n_{\eta_Q} \, ,
\label{eq:relationk2n}
\end{equation}
we can sum up the relativistic corrections to
\begin{equation}
      {\cal F}_{\eta_Q}(t_1,t_2) = \frac{  \bra{\eta_Q} | \psi^{\dag} \chi | \ket{0} }{2 \sqrt{2N_c}} \, \frac{ {\cal F}_{^1S_0}(t_1,t_2,\langle \vert \vec k \vert^2 \rangle_{\eta_Q})}{E(\langle \vert \vec k \vert^2 \rangle _{\eta_Q})} \, .
\end{equation}

It is convenient to represent this result as
\begin{equation}
    {\cal F}_{\eta_Q}(t_1,t_2)  = {\cal F}_0(t_1,t_2) \, \Big\{ 1 + g(t_1,t_2, \langle v^2 \rangle_{\eta_Q}) \Big \} \, ,  
\end{equation}
where
\begin{equation}
 g(t_1,t_2, \langle v^2 \rangle_{\eta_Q}) =  \frac{  \bra{\eta_Q} | \psi^{\dag} \chi | \ket{0} }{2 \sqrt{2N_c}} \, \frac{ {\cal F}_{^1S_0}(t_1,t_2,\langle \vert \vec k \vert^2 \rangle_{\eta_Q})}{E(\langle \vert \vec k \vert^2 \rangle_{\eta_Q}) \, {\cal F}_0(t_1,t_2)} - 1  , 
\end{equation}
where we had defined ${\cal F}_0$ in eq.~(\ref{eq:TFF_LO}). Using the result for ${\cal F}_{^1S_0}$ in eq.~(\ref{eq:FF_sd_all_order}), we then obtain the expression quoted in the main text in eq.~(\ref{eq:g_t1t2}).

In the limit $t_1 \to 0$ and $t_2 \to 0$, relevant for the di-photon decay width of $\eta_Q$, this expression simplifies to
\begin{equation}
    g(0,0,\langle v^2 \rangle_{\eta_Q}) = \frac{1}{2 \sqrt{\langle v^2 \rangle_{\eta_Q}} \left(\sqrt{1+ \langle v^2 \rangle_{\eta_Q}}\right)^3} \log \Bigg[ \frac{ \sqrt{1 + \langle v^2 \rangle_{\eta_Q}} + \sqrt{\langle v^2 \rangle_{\eta_Q}} }{\sqrt{1 + \langle v^2 \rangle_{\eta_Q}} - \sqrt{\langle v^2 \rangle_{\eta_Q}}} \Bigg] -1.
\label{eq:allorderrelativistict10t20}
\end{equation}

\section{Analytical expression for $K_1$}
\label{sec:appendixanalyticalK1}

In this appendix, we present the analytical expression for the one-loop coefficient $a_F$ in $K_1$. The renormalised amplitude contains functions in the class of multiple polylogarithms (MPLs). In the following, we define the function $G$, which can be written as iterated integrals~\cite{Goncharov:1998kja,Goncharov:2001iea} of the form
\begin{equation}
\begin{split}
    G{\left(a_1,...,a_n;x\right)}=&\int_0^x \frac{dt}{t-a_1}G{\left(a_2,...,a_n;t\right)},
\end{split}
\end{equation}
where the recursion ends with
\begin{equation}
    G{\left(;x\right)}=1.
\end{equation}
In the special case where the MPL has $a_n=0$, we need to use the alternative definition
\begin{equation}
G{\left(\vec{0}_k;x\right)}=\frac{1}{k!}\log^k{x},
\end{equation}
where $\vec{0}_k$ is a vector of zeroes of length $k$. The number of indices $n$ in the $G$ function indicates the weight of the multiple polylogarithm. The renormalised amplitude contains MPLs up to weight 2. These can be evaluated numerically using the $\mathtt{PolyLogTools}$ package \cite{Duhr:2019tlz} via its interface to $\mathtt{GiNaC}$ \cite{Vollinga:2004sn}.

We can express $a_F$ as a function of $t_1$ and $t_2$ in a compact way as
\begin{align}
   & a_F{(t_1, t_2)} = \frac{1}{2 r_3 \sqrt{r_3}} G{\left(0;-\frac{2}{t_1+t_2-4}\right)} \Bigg[p_1 G{\left(\frac{-\sqrt{r_3}+3 t_1+t_2-4}{4 t_1};1\right)}
   \nonumber \\
   &-p_1 G{\left(\frac{\sqrt{r_3}+3 t_1+t_2-4}{4 t_1};1\right)}-p_2 G{\left(\frac{-\sqrt{r_3}+t_1+3 t_2-4}{4 t_2};1\right)}
   \nonumber \\
   &+p_2 G{\left(\frac{\sqrt{r_3}+t_1+3 t_2-4}{4 t_2};1\right)}\Bigg] + \frac{p_1}{2 r_3 \sqrt{r_3}} \Bigg[ G{\left(\frac{-\sqrt{r_3}+3 t_1+t_2-4}{4 t_1},\frac{\sqrt{r_1} +1}{2};1\right)}
   \nonumber \\
   &+G{\left(\frac{-\sqrt{r_3}+3 t_1+t_2-4}{4 t_1},\frac{1-\sqrt{r_1}}{2};1\right)}-G{\left(\frac{\sqrt{r_3}+3 t_1+t_2-4}{4 t_1},\frac{\sqrt{r_1} +1}{2};1\right)}
   \nonumber \\
   &-G{\left(\frac{\sqrt{r_3}+3 t_1+t_2-4}{4 t_1},\frac{1-\sqrt{r_1}}{2};1\right)}-G{\left(\frac{-\sqrt{r_3}+3 t_1+t_2-4}{4 t_1},1;1\right)}
   \nonumber \\
   &+G{\left(\frac{\sqrt{r_3}+3 t_1+t_2-4}{4 t_1},1;1\right)} \Bigg] + \frac{p_2}{2 r_3 \sqrt{r_3}} \Bigg[ G{\left(\frac{\sqrt{r_3}+t_1+3 t_2-4}{4 t_2},\frac{\sqrt{r_2} +1}{2};1\right)}
   \nonumber \\
   &+G{\left(\frac{\sqrt{r_3}+t_1+3 t_2-4}{4 t_2},\frac{1-\sqrt{r_2}}{2};1\right)}-G{\left(\frac{-\sqrt{r_3}+t_1+3 t_2-4}{4 t_2},\frac{\sqrt{r_2}+1}{2};1\right)}
   \nonumber \\
   &-G{\left(\frac{-\sqrt{r_3}+t_1+3 t_2-4}{4 t_2},\frac{1-\sqrt{r_2}}{2};1\right)}+G{\left(\frac{-\sqrt{r_3}+t_1+3 t_2-4}{4 t_2},1;1\right)}
   \nonumber \\
   &-G{\left(\frac{\sqrt{r_3}+t_1+3 t_2-4}{4 t_2},1;1\right)} \Bigg]+\frac{\sqrt{r_1} t_1 \left(8 t_1 t_2-3 (t_1-4)^2+3 t_2^2\right)}{2 r_3 (t_1+t_2-4)}G{\left(0;\frac{\sqrt{r_1}-1}{\sqrt{r_1}+1}\right)}
   \nonumber \\
   &+\frac{\sqrt{r_2} t_2 \left(3 t_1^2+8 t_1 t_2-3 (t_2-4)^2\right)}{2 r_3 (t_1+t_2-4)}G{\left(0;\frac{\sqrt{r_2}-1}{\sqrt{r_2}+1}\right)}+\frac{1}{2 (t_1+t_2-2)}-\frac{9}{4}
   \nonumber \\
   &+\left(\frac{4 t_1 t_2}{r_3}-\frac{1}{t_1+t_2-2}+\frac{1}{(t_1+t_2-2)^2}-\frac{3}{4}\right) G{\left(0;-\frac{t_1}{2}-\frac{t_2}{2}+2\right)},
\label{eq:K1_factor}
\end{align}
where we have defined the roots, $r_i$, as
\begin{equation}
\begin{split}
    r_1 =& \frac{t_1-4}{t_1}, \qquad\qquad    r_2 = \frac{t_2-4}{t_2},
    \\
    r_3 =& t_1^2-2 t_1 t_2-8 t_1+t_2^2-8 t_2+16,
\end{split}
\end{equation}
and the polynomials, $p_i$, as
\begin{equation}
\begin{split}
    p_1 =& \,\tilde{p}{\left(t_1, t_2\right)}, \qquad\qquad    p_2 = -\tilde{p}{\left(t_2, t_1\right)},
    \\
    \tilde{p}{\left(a, b\right)}=&\, 16 (b-2)-(a-b) \left(a^2-3 a (b+2)-2 b\right).
\end{split}
\end{equation}

The result is applicable to both time-like and space-like regions. In order to perform the analytical continuation to time-like regions, one has to assign the quantities $t_1$ and $t_2$ with positive imaginary part:
\begin{equation}
    t_1 \rightarrow t_1 +i \delta,\qquad\qquad t_2 \rightarrow t_2 +i \delta.
\end{equation}

In the on-shell limit for both photons, $t_1=t_2=0$, the coefficient becomes
\begin{equation}
a_F\left(0,0\right)=\frac{\pi^2}{8}-\frac{5}{2}
\end{equation}
and reproduces the known result for $\gamma \gamma \leftrightarrow {^1S_0}$ \cite{Czarnecki:2001zc}. In the limit when only one photon becomes on-shell, $t_{1,2}=0$, we have checked that it reproduces the known results in the literature in both time-like and space-like regions \cite{Feng:2015uha,Sang:2009jc}. For the time-like region, the $a_F$ factor contains both a real and an imaginary part. For the NLO calculation however, only the real part is needed. In ref.~\cite{Sang:2009jc}, the imaginary part has been omitted. In contrast to this, our result also includes the imaginary part which can be used for the NNLO calculation in the time-like region.

\bibliography{references.bib}
\end{document}